\documentclass[prb,superscriptaddress]{revtex4}

\usepackage{epsfig}
\usepackage{bm}
\begin{document}  

\title{Multiple temperature scales of the periodic Anderson model:\\ 
the slave bosons approach}  
\author{S. Burdin} 
\affiliation{ 
Institut f\"ur Theoretische Physik, 
Universit\"at zu K\"oln, 
Z\"ulpicher Str. 77, 50937 K\"oln, Germany}
\author{V. Zlati\'c}  
\affiliation{ 
Institute of Physics, P.O. Box 304, 10001 Zagreb, Croatia}  
\date{\today}  
\begin{abstract} 
The thermodynamic and transport properties of intermetallic 
compounds with Ce, Eu, and Yb ions are discussed using the 
periodic Anderson model with an infinite correlation between 
$f$ electrons. 
At high temperatures, these systems exhibit typical features 
that can be understood in terms of a single impurity Anderson or 
Kondo model with Kondo scale T$_K$. 
At low temperatures, the normal state is governed by the Fermi 
liquid (FL) laws with characteristic energy scale T$_0.$ 
The slave boson solution of the periodic model shows that T$_0$ 
and  T$_K$ depend not only on the degeneracy and the splitting 
of the $f$ states, the number of $c$ and $f$ electrons, and their coupling, 
but also on the shape of the conduction electrons density of states 
($c$ DOS) in the vicinity of the chemical potential. 

We show that the details of the band structure determine the 
ratio T$_0$/T$_K$ and that the crossover between the high- and 
low-temperature regimes in ordered compounds is system-dependent. 
A sharp peak in the $c$ DOS yields T$_0 \ll$T$_K$ and explains 
the 'slow crossover' observed in YbAl$_3$ or YbMgCu$_4$.
A minimum in the $c$ DOS yields T$_0 \gg$T$_K$, which leads to 
the abrupt transition between the high- and low-temperature 
regimes in YbInCu$_4$. 
In the case of CeCu$_2$Ge$_2$ and CeCu$_2$Si$_2$, where 
T$_0 \simeq T_K$, the slave boson solution explains the pressure 
experiments which reveal sharp peaks in the T$^2$ coefficient 
of the electrical resistance, $A=\rho(T)/T^2$, and the residual 
resistance. These peaks are due to the change in the degeneracy 
of the $f$ states induced by the applied pressure. 

The FL laws explain also the correlation between the specific 
heat coefficient $\gamma=C_V/T$ and the slope of the thermopower 
$\alpha(T)/T$, or between $\gamma$ and the $A$ coefficient 
of the resistivity.  
For $N$-fold degenerate model, the FL laws explain the 
deviations from universal value of the Kadowaki-Woods ratio, 
$R_{KW}=A/\gamma^2$, and the ratio $q=lim_{\{T\to 0\}}\alpha/\gamma T$.
The renormalization of transport coefficients can invalidate 
the Wiedemann-Franz law and lead to an enhancement of the 
thermoelectric figure-of-merit. 
We show that the low-temperature response of the periodic 
Anderson model can be enhanced (or reduced) with respect to the 
predictions based on the single-impurity models that give the 
same high-temperature behavior.
\end{abstract} 
\maketitle

 
\section{Introduction} 
Intermetallic compounds containing Cerium, Ytterbium and Europium  
ions exhibit a number of remarkable phenomena, like the heavy fermion 
mass enhancement, valence fluctuations, huge thermopower, 
spin-charge separation, unconventional magnetism and superconductivity, etc.  
These systems have been studied for several decades 
and are still attracting considerable attention. 
The initial focus was on dilute alloys with magnetic 3$d$ 
and 4$f$ impurities, where the low- and high-temperature behavior 
is qualitatively different. 
At high temperatures, the experimental data indicate that 
the $f$ states are localized and weakly coupled to conduction states.  
The susceptibility is of the Curie-Weiss form and entropy is large, 
as expected of nearly free $f$ ions. At the same time, 
the logarithmic resistivity and large thermopower are typical of 
the conduction ($c$) electrons weakly perturbed by the local moments. 
Such a behavior is explained by the perturbative solution of the 
Anderson or Kondo models, which yield the low-energy  correlation 
functions as universal functions of reduced temperature T/T$_K$, 
where T$_K$ is the Kondo temperature. 
At  low temperatures, the experimental data show that 
the susceptibility is Pauli like, the specific heat and 
entropy are linear in temperature, and transport coefficients 
are given by simple powers of reduced temperature T/T$_K$.  
Close to the ground state, the $c$ and $f$ states seem to 
form a non-magnetic singlet with the fermionic excitation 
spectrum and energy scale T$_K$. 
The overall behavior of dilute alloys with paramagnetic 
impurities is nearly the same when plotted on the $T/T_K$ scale, 
even though the values of T$_K$ vary by orders of magnitude.  
Unlike the high-temperature data, neither the crossover 
nor the low-temperature behavior can be understood in terms of the 
perturbation theory which treats the conduction electrons and 
the local moments as separate entities.  
The fact that the low-temperature scale coincides with the  
high-temperature one and that the crossover from the weak 
to the strong coupling regime takes place around T$_K$  
are the most prominent features of dilute Kondo systems.  
 
The dilute alloy problem has been explained after several decades 
of intensive work by the exact solutions of the Kondo and 
the Anderson models.  
Early calculations considered the models in which the conduction 
electrons with a constant density of states ($c$ DOS)    
are exchange scattered on a spin-1/2 local moment.  
The solution was obtained by the variety of methods, like 
the perturbative scaling\cite{anderson.1970},  
the numerical renormalization group\cite{wilson.1975}, 
the Fermi liquid theory\cite{Nozieres.1974,yamada.75a},  
and Bethe Ansatz\cite{tsvelick.1983}. 
These results reveal that the effective coupling between 
the free fermions and degenerate local moments increases at 
low temperatures and diverges in the ground state, which 
explains the breakdown of the perturbation theory. 
However, the simple models cannot provide a quantitative 
description of dilute Kondo alloys; a realistic modeling 
has to take into account the details of the local states 
and the band structure of the host, and consider additional 
scattering mechanisms. 
In Kondo systems with Ce and Yb ions, the splitting of the 4$f$ 
states by the spin-orbit and the crystal field (CF) effects 
can change the effective degeneracy of the system, reduce 
the Kondo scale, and lead to the seemingly complicated features, 
like the multiple peaks in the resistivity or the sign change 
of the thermopower and the Hall effect.  
A non-constant $c$ DOS leads also to a complicated low-temperature 
behavior. For example, Whithoff and Fradkin~\cite{wf.1990} 
have found that the $c$ DOS with a power-law singularity leads  
to a critical coupling $J_{c}$ that separates different ground states. 
For $J>J_{c}$, the ground state is of the strong coupling type, 
while for $J<J_{c}$ the renormalized coupling decreases with 
temperature and the ``usual'' Kondo screening of local moments does 
not occur\cite{vojta.2002}. Nevertheless, despite a conduction
electron pseudo-gap, it has been shown that Kondo-lattice screening 
can be more stable than in the single Kondo impurity 
case~\cite{schmalian.2008}. 

The Kondo problem becomes much more difficult for stoichiometric 
compounds in which a magnetic ion is present in each unit cell.  
The high-temperature features can still be explained by an 
effective single-impurity model, if one takes into account 
additional splittings of the $f$ states and/or the fact that 
$c$ DOS can change rapidly in the vicinity of the chemical potential.  
The perturbation theory provides a consistent picture of 
the high-temperature data: 
it yields the correct Kondo scale, explains the Curie-Weiss 
behavior of the susceptibility and the logarithmic decrease 
of the resistivity and thermopower, and accounts for the well 
resolved CF excitations seen in neutron experiments.
However, at sufficiently low temperatures the scattering becomes 
coherent and one finds new features that cannot be explained 
by the single impurity models.  
The onset of coherence is most clearly seen in the electrical resistivity 
which drops to very small values, and in the optical conductivity 
which shows the development of a low-frequency Drude peak and a small 
hybridization gap close to the chemical potential.   
At lowest  temperatures, the Fermi liquid laws often emerge: 
the resistivity is quadratic and the thermopower a linear function 
of temperature; 
the specific heat and the magnetic susceptibility are much enhanced,  
indicating a large effective electronic mass;  
the de Haas-van Alphen experiments show that $f$ electrons 
contribute to the Fermi volume.
The low-temperature ratios of various correlation functions, 
like the Wilson ratio, $\chi/\gamma$ or the Kadowaki-Woods ratio, 
$A/\gamma^2$,  assume  the universal values. 
Here, $\gamma$ and $\chi$, denote the $T\to 0$ limit of 
the specific heat coefficient and the magnetic susceptibility, 
and $A$ is the coefficient of the $T^2$ term in the resistivity. 
The universality of these ratios indicates that the ground state 
properties depend on a single energy scale T$_0$. 
However, unlike in dilute alloys, this FL scale T$_0$ 
can be much different from T$_K$. 

In this paper we study the periodic systems with 4$f$ ions 
and show that the relative magnitude of the Kondo  and the 
FL scale depends on the shape of the unperturbed 
conduction states.  
The Anderson lattice with an enhanced DOS around the chemical 
potential has T$_0\ll$T$_K$, which explains the gradual transition 
between the coherent and the incoherent regimes (slow crossover) 
observed in YbAl$_3$~\cite{bauer:125102}.  
A pseudo-gap or a reduced $c$ DOS close to the  chemical potential 
yields T$_K\ll$T$_0$, which explains the abrupt valence-change 
transition observed in Yb- and Eu-based intermetallic compounds,  
like  YbInCu$_4$~\cite{sarrao.1999},  
EuNi$_2$(Si$_{1-x}$Ge$_{x}$)$_2$~\cite{wada.1997} or  
Eu(Pd$_{1-x}$Pt$_{x}$)$_2$Si$_2$~\cite{mitsuda.2000}.  
In the case of CeCu$_2$Ge$_2$ and CeCu$_2$Si$_2$, where 
T$_K$ and T$_0$ seem to be of the same order of magnitude, 
we show that the pressure-dependence of the $A(P)$ coefficient 
and the residual resistance is driven by the change in the 
degeneracy of the $f$ states. 

The  paper is organized as follows. Section ~\ref{formalism} 
explains briefly the slave boson formalism for the 
Anderson model and defines its low- and high-temperature 
scales T$_0$ and T$_K$. 
This section provides also the relationship between 
the $c$ DOS and the relative magnitude of T$_0$ and T$_K$, 
discusses the effects of the magnetic field, and shows how 
to express the transport coefficients in terms of T/T$_0$. 
In Section ~\ref{results} we discuss the relevance of these results 
for the experimental data on various intermetallic compounds in which 
the ratio T$_0/$T$_K$ can be much smaller or larger than one. 
For the case T$_0\simeq$T$_K$, we discuss the pressure anomalies 
in the $A$ coefficient and the residual resistance induced 
by the changes in the effective degeneracy of the $f$ states. 

\section{Slave boson solution of the periodic Anderson model\label{formalism}} 

\subsection{The mean-field self consistent equations in the limit  $U\to\infty$ } 

\begin{figure}[h] 
\includegraphics[width=3.0cm,angle=-90]{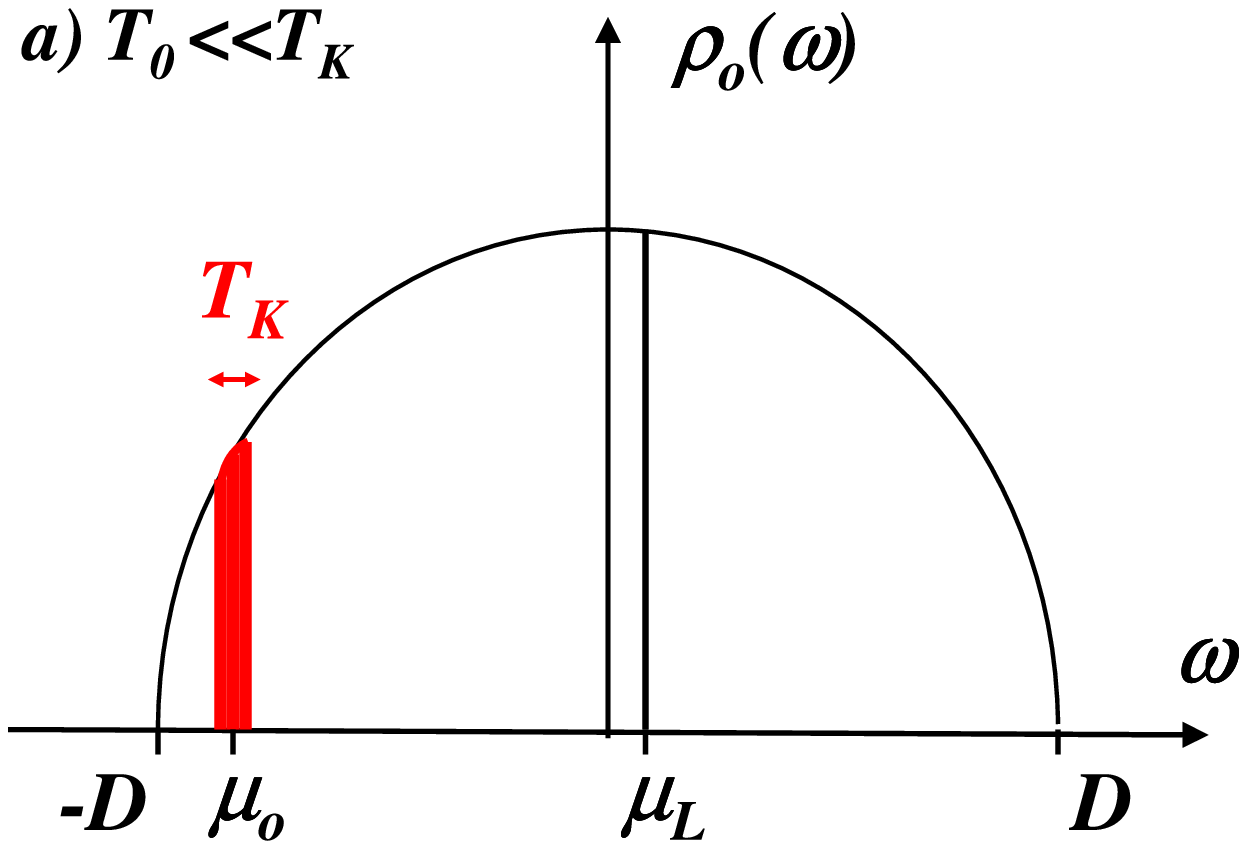} 
\includegraphics[width=3.0cm,angle=-90]{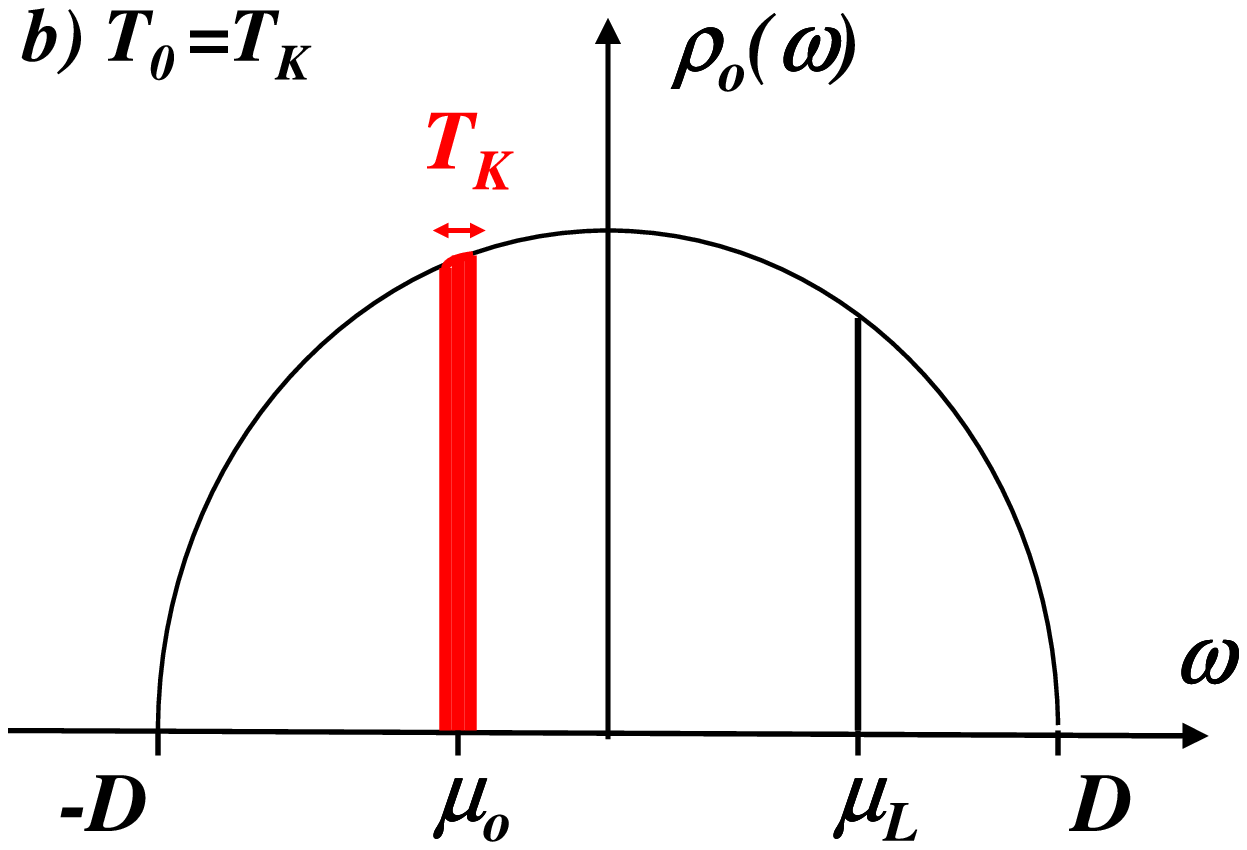} 
\includegraphics[width=3.0cm,angle=-90]{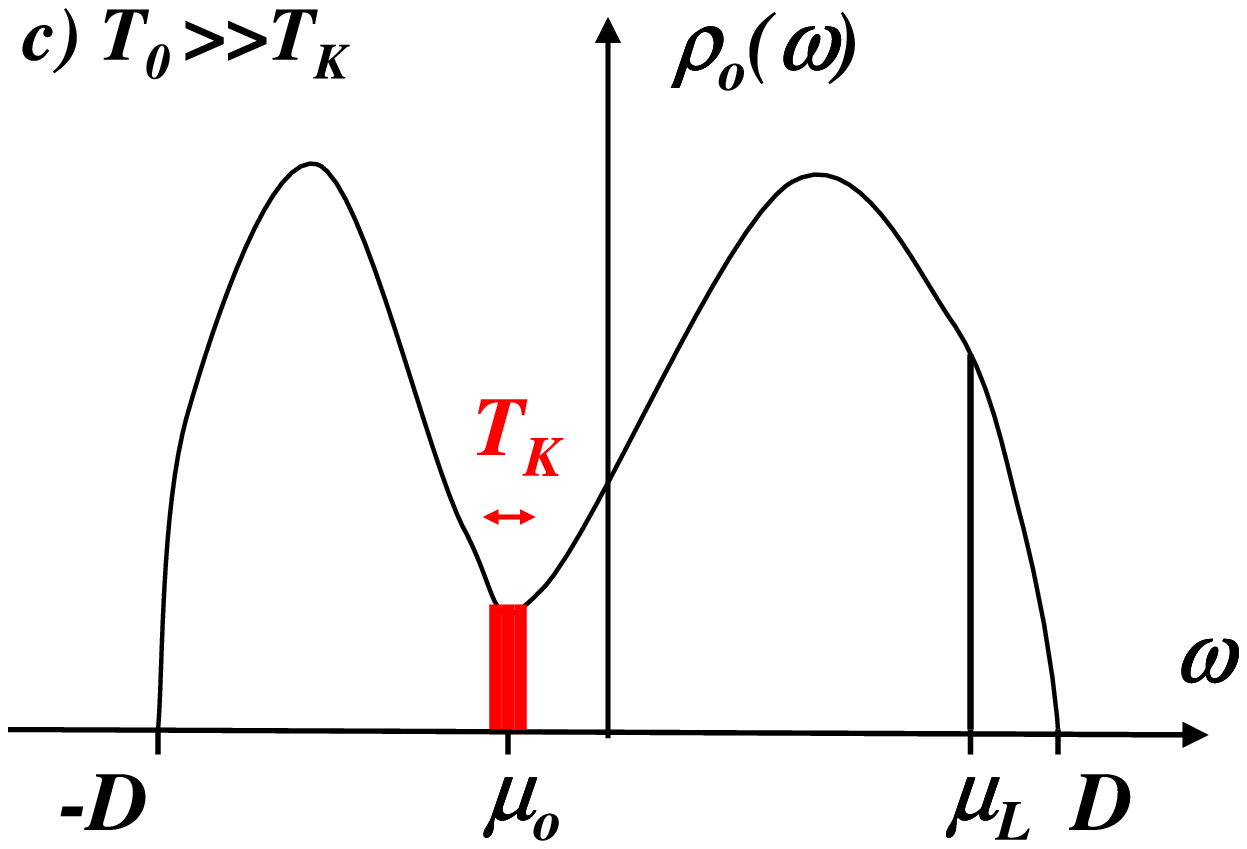}   
\caption{(Color online) Schematic plot of the non-interacting 
$c$ DOS $\rho_{0}(\omega)$. 
(a) For a regular $c$ DOS and far from the electronic half-filling 
we find T$_{0}\ll$T$_{K}$.  
Close to the half-filling, T$_{0}/$T$_{K}$ depends on the shape 
of the $c$ DOS around the renormalized chemical potential $\mu$: 
(b) $\rho_{0}(\omega)$ is nearly constant for $\omega\sim \mu$ 
and T$_{0}\sim$T$_{K}$. 
(c) $\mu$ is close to a minimum of $\rho_{0}(\omega)$ and 
T$_{0}\gg$T$_{K}$.  
Here, $\mu_{0}\approx\mu$ is the chemical potential of $n_{c}=n-1$ 
non-interacting $c$ electrons (``small Fermi surface''), and 
$\mu_{L}$ is the chemical potential of $n_{c}=n$ non-interacting 
$c$ electrons (``large Fermi surface''). 
} 
\label{FigSchemasDOS} 
\end{figure}

The periodic Anderson Model (PAM) Hamiltonian is written 
in the limit of an infinite correlation between $f$ electrons, 
$U=+\infty$, as 
\begin{eqnarray} 
H
=&&
\sum_{{\bf k}\sigma} 
\epsilon_{\bf k}c_{{\bf k}\sigma}^{\dagger}c_{{\bf k}\sigma} 
+E_{f}\sum_{i\sigma}f_{i\sigma}^{\dagger}f_{i\sigma} 
+V\sum_{i\sigma}[c_{i\sigma}^{\dagger}f_{i\sigma} 
+f_{i\sigma}^{\dagger}c_{i\sigma}] 
-\mu\sum_{i\sigma}[f_{i\sigma}^{\dagger}f_{i\sigma} 
+c_{i\sigma}^{\dagger}c_{i\sigma}] \cr
&+&
h\sum_{i}
\left[  
g_{c}(c_{i\uparrow}^{\dagger}c_{i\uparrow} 
-c_{i\downarrow}^{\dagger}c_{i\downarrow}) 
+ 
g_{f}(f_{i\uparrow}^{\dagger}f_{i\uparrow} 
-f_{i\downarrow}^{\dagger}f_{i\downarrow}) 
\right]~,  
\label{PAM} 
\end{eqnarray} 
where $c_{i\sigma}$ and $f_{i\sigma}$ are annihilation operators  
for $c$ and $f$ electrons, $i$ is the site index, 
$\sigma=\uparrow,\downarrow$ is the spin component. 
$h$ is an external magnetic field,  
and $g_{c}$ and $g_{f}$ the Land\'e factors for $c$ and $f$ 
orbitals,  respectively. 
The $c$ orbital describes the conduction band with energy levels  
$\epsilon_{\bf k}$, where ${\bf k}$ is the momentum component in 
the reciprocal space, the localized $f$ orbitals are characterized 
by energy level $E_{f}$ and the local hybridization between the two orbitals 
is specified  by the matrix element $V$. 
The infinite Coulomb repulsion restricts the occupation of 
the $f$ states to $n_f\leq 1$ and we use  the chemical potential 
$\mu$ to fix the total electronic occupation per site to $n=n_c+n_f$. 
The unperturbed $c$ DOS is $\rho_{0}(\omega)=
{1}/{\cal N} \sum_{\bf k}\delta(\omega-\epsilon_{\bf k})$, 
where ${\cal N}$ is the number of lattice sites. 
In the following, $\rho_{0}$ will be characterized by a half bandwidth 
$D$. 
Typical band-shapes and fillings considered in this work are 
shown in Fig.~\ref{FigSchemasDOS}. All the energies (except the 
excitation energies) are measured with respect to the center  
of the $c$-band (see Fig.~\ref{FigSchemasDOS}).  
The effective degeneracy of the model is determined by the lowest 
spin-orbit state or the additional crystal field splitting of 
the $f$ states of the Ce and Yb ions. 
This degeneracy can be changed by temperature, pressure or 
chemical pressure. Here, we consider an effective spin-1/2 model.  

The model defined by Eq.(\ref{PAM}) is solved for an arbitrary $c$ DOS 
by the slave boson  mean-field (MF) 
approximation~\cite{Slaveboson1,Slaveboson2}, 
which represents the $f$ states by the product of spinless bosons,
$b_{i}^{\dagger}$, and auxiliary fermions of 
spin $\sigma=\uparrow,\downarrow$, $d_{i\sigma}^{\dagger}$. 
By definition, these auxiliary particle operators 
create the local $f$ states with no electron and one $\sigma$-spin electron,
respectively;   
$\left| 0 \right>_{i}^{f}\longrightarrow  
b_{i}^{\dagger}\left| 0 \right>_{i}$ and 
$\left| \sigma \right>_{i}^{f} \longrightarrow
d_{i\sigma}^{\dagger}\left| 0 \right>_{i}$. 
The electronic operators are related to the auxiliary operators as 
$f_{i\sigma}=b_{i}^{\dagger}d_{i\sigma}$, which leads to the local identities 
$f_{i\sigma}^{\dagger}f_{i\sigma}=d_{i\sigma}^{\dagger}d_{i\sigma}$.   
The doubly occupied state $\left| \uparrow\downarrow \right>_{i}^{f}$ is  
forbidden in the limit $U\to\infty$.
The anti-commutation relations for $f$ operators as well as the physical  
Hilbert space are recovered by enforcing the local constraints  
$b_{i}^{\dagger}b_{i}+\sum_{\sigma}d_{i\sigma}^{\dagger}d_{i\sigma}=1$.  
The latter are satisfied by introducing the time-dependent Lagrange 
multipliers  $\lambda_{i}(\tau)$.  
Assuming $g_{c}<<g_{f}$ we set $g_{c}=0$ and $g_{f}=1$, 
replace $hg_{f}\rightarrow h$, and rewrite the PAM  Hamiltonian~(\ref{PAM}) 
in terms of auxiliary fermionic and bosonic operators,
\begin{eqnarray} 
H_{SB}&=&\sum_{{\bf k} \sigma} 
\epsilon_{\bf k}c_{{\bf k}\sigma}^{\dagger}c_{{\bf k}\sigma} 
+V\sum_{i\sigma}[b_{i}^{\dagger}c_{i\sigma}^{\dagger}d_{i\sigma} 
+b_{i}d_{i\sigma}^{\dagger}c_{i\sigma}] 
-\mu\sum_{i\sigma}c_{i\sigma}^{\dagger}c_{i\sigma} 
+h\sum_{i}(d_{i\uparrow}^{\dagger}d_{i\uparrow} 
-d_{i\downarrow}^{\dagger}d_{i\downarrow}) \nonumber \\
&&+(E_{f}-\mu)\sum_{i}(1-b_{i}^{\dagger}b_{i}) 
+\sum_{i} \lambda_{i}
\left[ { 
b_{i}^{\dagger}b_{i}-1+\sum_{\sigma}d_{i\sigma}^{\dagger}d_{i\sigma} }
\right]~. 
\label{PAMbosons} 
\end{eqnarray} 
Since the slave boson Hamiltonian~(\ref{PAMbosons}) is invariant 
under the local gauge transformation $b_{i}\to b_{i}e^{i\theta_{i}}$,  
$d_{i\sigma}\to d_{i\sigma}e^{i\theta_{i}}$, we choose the gauge such that 
the bosonic fields are real, $b_{i}=b_{i}^{\dagger}\equiv r_{i}$. 
Finally, we make the MF approximation, assuming that the boson fields 
and  the Lagrange multipliers are homogeneous and spin-independent 
constants, $r_{i}\equiv r$ and  $\lambda_{i}\equiv \lambda$.
Within this MF approach, which is exact in the limit of  
a large number of spin components, the Hamiltonian~(\ref{PAMbosons})  
becomes quadratic, 
\begin{eqnarray} 
H_{SB}^{MF}=\sum_{{\bf k} \sigma}
\left[ 
{ \epsilon_{\bf k}c_{{\bf k}\sigma}^{\dagger}c_{{\bf k}\sigma} 
+rV\left( 
{ c_{{\bf k}\sigma}^{\dagger}d_{{\bf k}\sigma} 
+d_{{\bf k}\sigma}^{\dagger}c_{{\bf k}\sigma} }
\right) 
-\mu c_{{\bf k}\sigma}^{\dagger}c_{{\bf k}\sigma} 
+ 
\lambda_{\sigma} 
d_{{\bf k}\sigma}^{\dagger}d_{{\bf k}\sigma} 
+\frac{E_{f}-\lambda-\mu}{2}(1-r^{2}) }
\right]~,  
\label{PAMmagnetbosonsMF} 
\end{eqnarray} 
where 
\begin{eqnarray} 
\lambda_{\uparrow}&\equiv&\lambda +h~, \\ 
\lambda_{\downarrow}&\equiv&\lambda -h~.  
\end{eqnarray} 
In the presence of the magnetic field $h$, 
the Lagrange multipliers are shifted by $\pm h$ 
but the up- and down-spin states remain decoupled.
The self-consistent solution is obtained by minimizing 
the free energy  
$\beta {\cal F}(r, \lambda)\equiv -\ln Tr e^{-\beta H}$
with respect to $r$ and $\lambda$.  
>From ${\partial {\cal F}(r, \lambda)}/{\partial r}=0$ 
and ${\partial {\cal F}(r, \lambda)}/{\partial \lambda}=0$ 
we obtain 
\begin{eqnarray} 
2r (E_{f}-\lambda-\mu) 
&=& 
\frac{V}{\cal N}
\sum_{{\bf k}\sigma}\langle c_{{\bf k}\sigma}^{\dagger}d_{{\bf k}\sigma} 
+d_{{\bf k}\sigma}^{\dagger}c_{{\bf k}\sigma}\rangle~,   
\label{MFmagneteqr1}\\ 
r^{2} 
&=& 
1-
\frac{1}{\cal N}
\sum_{{\bf k}\sigma} 
\langle d_{{\bf k}\sigma}^{\dagger}d_{{\bf k}\sigma}\rangle~,  
\label{MFmagneteqlambda1} 
\end{eqnarray} 
where $\langle\cdots\rangle$ is the thermal average with 
respect to the MF Hamiltonian~(\ref{PAMmagnetbosonsMF}).  
The total electron occupation per site is   
\begin{eqnarray}  
n=\frac{1}{\cal N}\sum_{i\sigma}\langle d_{i\sigma}^{\dagger}d_{i\sigma} 
+c_{i\sigma}^{\dagger}c_{i\sigma}\rangle 
\equiv n_{f}+n_{c}~.   
\label{MFmagneteqmu1} 
\end{eqnarray} 
Here, $n_{f}\equiv 1/{\cal N} \sum_{i\sigma}\langle f_{i\sigma}^{\dagger}f_{i\sigma}\rangle 
= 1/{\cal N}\sum_{i\sigma}\langle d_{i\sigma}^{\dagger}d_{i\sigma}\rangle $  
and  
$n_{c}\equiv 1/{\cal N} \sum_{i\sigma}\langle c_{i\sigma}^{\dagger}c_{i\sigma}\rangle$,   
are the average occupations per site for $f$ and $c$ orbitals, respectively.  
Using the relationship between the slave-boson amplitude and  
the total electronic occupation,   
\begin{eqnarray} 
n_{f}&=&1-r^{2}~, \\ 
n_{c}&=&n-n_{f}~.  
\label{eq: magnetnf} 
\end{eqnarray} 
the self-consistency equations~(\ref{MFmagneteqr1}), 
(\ref{MFmagneteqlambda1}), and~(\ref{MFmagneteqmu1}) 
can be written for $r\neq 0$ as, 
\begin{eqnarray} 
\frac{\lambda+\mu-E_{f}}{V^2} 
&=& 
\frac{1}{rV} 
 \sum_\sigma
\int_{-\infty}^{+\infty} 
\rho_{dc}^{\sigma}(\omega) 
n_{F}(\omega)d\omega 
\label{MFmagneteqr2}~, 
\\ 
n_{f} 
&=& 
\sum_\sigma
\int_{-\infty}^{+\infty} 
\rho_{d}^{\sigma}(\omega) 
n_{F}(\omega)d\omega 
\label{MFmagneteqlambda2}
~, \\ 
n_{c} 
&=& 
\sum_\sigma
\int_{-\infty}^{+\infty} 
\rho_{c}^{\sigma}(\omega) 
n_{F}(\omega)d\omega 
\label{MFmagneteqmu2}~,  
\end{eqnarray} 
where $n_{F}(\omega)\equiv 1/[1+e^{\beta\omega}]$ is the Fermi function  
and $\rho_{dc}^{\sigma}$, $\rho_{d}^{\sigma}$, $\rho_{c}^{\sigma}$ 
the spectral densities of  the local single-particle Green's 
functions of a given spin component. 
The Green's functions are defined in the usual way as   
thermal averages of the (imaginary) time-ordered products of the  
appropriate creation and annihilation operators.  
For the quadratic MF Hamiltonian~(\ref{PAMmagnetbosonsMF}), 
the  Green's functions for $r\neq 0$ 
are easily calculated by the equations of motion,    
which yields\cite{zlatic.07b} 
\begin{eqnarray} 
\rho_{dc}^{\sigma}(\omega)
\equiv 
-\frac{1}{\pi} {\mathrm Im} \ G_{dc}^{\sigma}(\omega) 
&=& 
\frac{rV}{\omega-\lambda_{\sigma}} \rho_{c}^{\sigma}(\omega)
~, 
\label{eq:rho_dc}\\ 
&&\nonumber\\ 
\rho_{d}^{\sigma}(\omega)
\equiv 
-\frac{1}{\pi} {\mathrm Im} \ G_{d}^{\sigma}(\omega) 
&=& 
\frac{r^{2}V^{2}}{(\omega-\lambda_{\sigma})^{2}} 
\rho_{c}^{\sigma}(\omega)~,  
\label{eq:rho_d}\\ 
&&\nonumber\\ 
\rho_{c}^{\sigma}(\omega)
\equiv 
-\frac{1}{\pi} {\mathrm Im} \ G_{c}^{\sigma}(\omega) 
&=& 
\rho_{0}\left({ 
\omega+\mu-\frac{r^{2}V^{2}}{\omega-\lambda_{\sigma}} 
}\right)~.
\label{eq:rho_c}  
\end{eqnarray} 
In what follows we analyze the MF slave boson solution 
and discuss the effects of the band-structure. 

\subsection{The high-temperature solution} 
At high enough temperatures, 
the self consistency equations ~(\ref{MFmagneteqr1})
-- 
(\ref{eq: magnetnf}),  
have only a trivial $r=\lambda=0$ solution, i.e.,  the effective 
MF hybridization in Eq.~(\ref{PAMmagnetbosonsMF}) vanishes, $rV=0$. 
This solution describes a system of decoupled $f$ and $c$ electrons.  
For $h=0$ and $E_{f} < \mu$, 
each lattice site is occupied by a single $f$ electron 
of spin $\sigma$ and there are $n_{c}=n-1$ conduction electrons 
(per site) with the $c$ DOS given by $\rho_0(\epsilon)$.  
The Fermi surface encloses $n_{c}=n-1$ points in the k-space, 
i.e., the $r=0$ state occupies a "small" Fermi volume that 
includes "light" $c$ electrons but not the $f$ states. 
The magnetic susceptibility is Curie-like, provided the 
direct  effect of $h$ on the conduction electrons is neglected.
(The  Pauli-like susceptibility of $c$ electrons is negligible with 
respect to the Curie contribution of the $f$ states.)  
As long as the $c$ states are highly degenerate, $T\ll D$,  
their entropy is much smaller than the entropy of the localized 
$f$ states, and the overall entropy per site 
is approximately given by $S=\ln 2$. 
In the presence of a large magnetic field, the degeneracy of the $f$  
states is lifted and the system acquires additional Zeeman energy.  
 
Even though the trivial  $r=\lambda=0$ solution does not provide 
a quantitative description of the PAM at high temperatures, it captures 
the essential qualitative  point: it represents the whole system 
in terms of two well-defined but separated sub-systems.  
A more realistic approach would take into account the small coupling 
between the $c$ and $f$ electrons and treat it as a perturbation.  
This would reduce the average $f$ occupation to $n_{f}<1$,  
give the Curie-Weiss rather than the Curie susceptibility,  
and obtain the transport properties from the scattering 
of "light" $c$ electrons on the $f$ ions. The resistivity 
calculated in such a way has logarithmic corrections to 
the high-temperature spin disorder limit. 
However, both the trivial slave boson solution and the perturbative 
one break down at sufficiently low temperatures. 

Remarkably, the Kondo scale T$_K$ defined by the high-temperature 
perturbation expansion agrees with the characteristic temperature 
at which the non-trivial solution of he slave boson equations emerges.
In what follows we analyze the non-trivial solution of  the 
self-consistent equations ~(\ref{MFmagneteqr1})
-- (\ref{eq: magnetnf}) for the spin-1/2 model and show that 
it captures the main features of the experimental data on 
Ce, Eu and Yb intermetallic compounds at low temperatures. 
The generalization to an arbitrary $SU(N)$ symmetric Anderson model 
or the model with the CF split $f$ states, is straightforward 
(see Appendix). 
 
\subsection{Kondo temperature T$_{K}$ \label{TK} } 
 
The non-trivial $r\neq 0$ solution of the slave boson equations 
is found below some critical temperature which defines the Kondo 
scale T$_K$ of the periodic Anderson model.  
For a given set of parameters and total occupation $n$, the Kondo scale  
is obtained from the $r, \lambda\to 0$ limit of the equations 
~(\ref{MFmagneteqr1}) 
-- (\ref{MFmagneteqmu1}).  
This gives, for $h=0$,   
\begin{eqnarray} 
\frac{2}{J(\mu)} 
&=& 
\int_{-\infty}^{+\infty} 
\frac{\rho_{0}(\omega+\mu)\tanh{(\omega/2T_{K})}}{\omega} 
d\omega~,  
\label{EqTK} \\ 
n_{f} 
&=& 
2\int_{-\infty}^{+\infty}d\omega
~\delta (\omega-\lambda)~ n_{F}(\omega)~=~1~, \\ 
n_c
&=& 
n-1
=
2\int_{-\infty}^{+\infty}d\omega 
~{\rho_{0}(\omega+\mu)} ~ n_{F}(\omega)~,  
             \label{EqTK_nc} 
\end{eqnarray} 
where we introduced the Kondo coupling constant,   
\begin{eqnarray} 
J(\mu) 
\equiv \frac{2V^{2}}{\mu-E_{f}}. 
\label{DefKondocoupling} 
\end{eqnarray} 
Assuming that T$_K$  vanishes continuously and 
taking the limit $T_K\to 0$ yields the critical coupling  
\begin{equation}  
\frac{2}{J_{c}}  
=  
\int_{-\infty}^{+\infty}d\omega  
\frac{\rho_0(\omega+\mu)}{|\omega |}.    
\label{DivergentCouplingT=0} 
\end{equation}  
For a regular DOS, $\rho_0(\mu)\neq 0$, the r.h.s. of 
Eq.(\ref{DivergentCouplingT=0}) diverges logarithmically, 
such that $J_{c}=0$ and any finite coupling leads to  $T_K >0$. 
At T$_K$, the high-temperature $r=0$ phase with large 
paramagnetic entropy is destabilized by a transition to 
the low-entropy Fermi liquid phase. The Kondo scale given 
by Eq.~(\ref{EqTK}) coincides with the solution of the scaling 
equations for the $SU(2)$ single-impurity Kondo model~\cite{wf.1990}.  

The physical interpretation of the emergence of the non-trivial slave 
boson solution is made by the analogy with the single-impurity case. 
We assume that the high-temperature solution of the lattice describes 
a system of localized $f$ and $c$ electrons with small FS,  
and that the correction to the $r=0$ solution can be found by 
the perturbation theory in terms of $J(\mu)$. 
For $T>T_{K}$ the hybridization energy is small with respect 
to the entropic contribution to the free energy due to the 
degenerate $f$ states. Thus, the total free energy of the 
system, ${\cal F}=E-TS$, is minimized at high temperatures 
by the paramagnetic configuration in which the $f$ states 
are very weakly coupled to the $c$ states. 
For $T\leq T_K$, the entropic contribution is reduced below 
the hybridization energy and the paramagnetic configuration 
becomes thermodynamically  unstable.  At T$_K$, the crossover 
between the local moment and a local singlet state takes place.  

In the case of an unperturbed $c$ DOS with a pseudo-gap at $\mu$, 
$\rho_0(\mu)=0$, equations~(\ref{MFmagneteqr2}) 
-- (\ref{MFmagneteqmu2}) 
yield the solution with finite $J_c$ which separates two 
different low-temperature regimes.  
For $J>J_c$ the $r\neq 0$ solution emerges at temperature T$_K$ 
which is the same as in the $\rho_0(\mu)\neq 0$ case. 
But for $J<J_c$  the $r=0$ solution persists down to $T=0$, such that    
the paramagnetic entropy is not removed by Kondo scattering.
If the coupling is tuned by pressure or doping, a quantum  
phase transition can be induced at the critical value $J_c$ (or $V_c$).  
In the case of $c$ DOS with a gap $\Delta_{g}\ll D$ around  
the Fermi level we find $J_{c}\propto D/ln[D/\Delta_{g}]$.   
A pseudo-gap centered at the chemical potential $\mu$  
and characterized by a single energy scale $D$ gives $J_{c}\propto D$.  
The power-law singularity $\rho_{0}(\omega)=R_{0} |\omega|^\eta$ gives  
$J_{c}=2\eta D/(\eta+1)$, which is the Whithoff and Fradkin  
result~\cite{wf.1990,vojta.2002}. (The constant $R_{0}$ follows from the  
normalization condition, $\int_{-D}^{D} \rho_0(\omega)= 1$.)
 
\begin{figure}[h] 
\includegraphics[width=6.0cm,angle=-90]{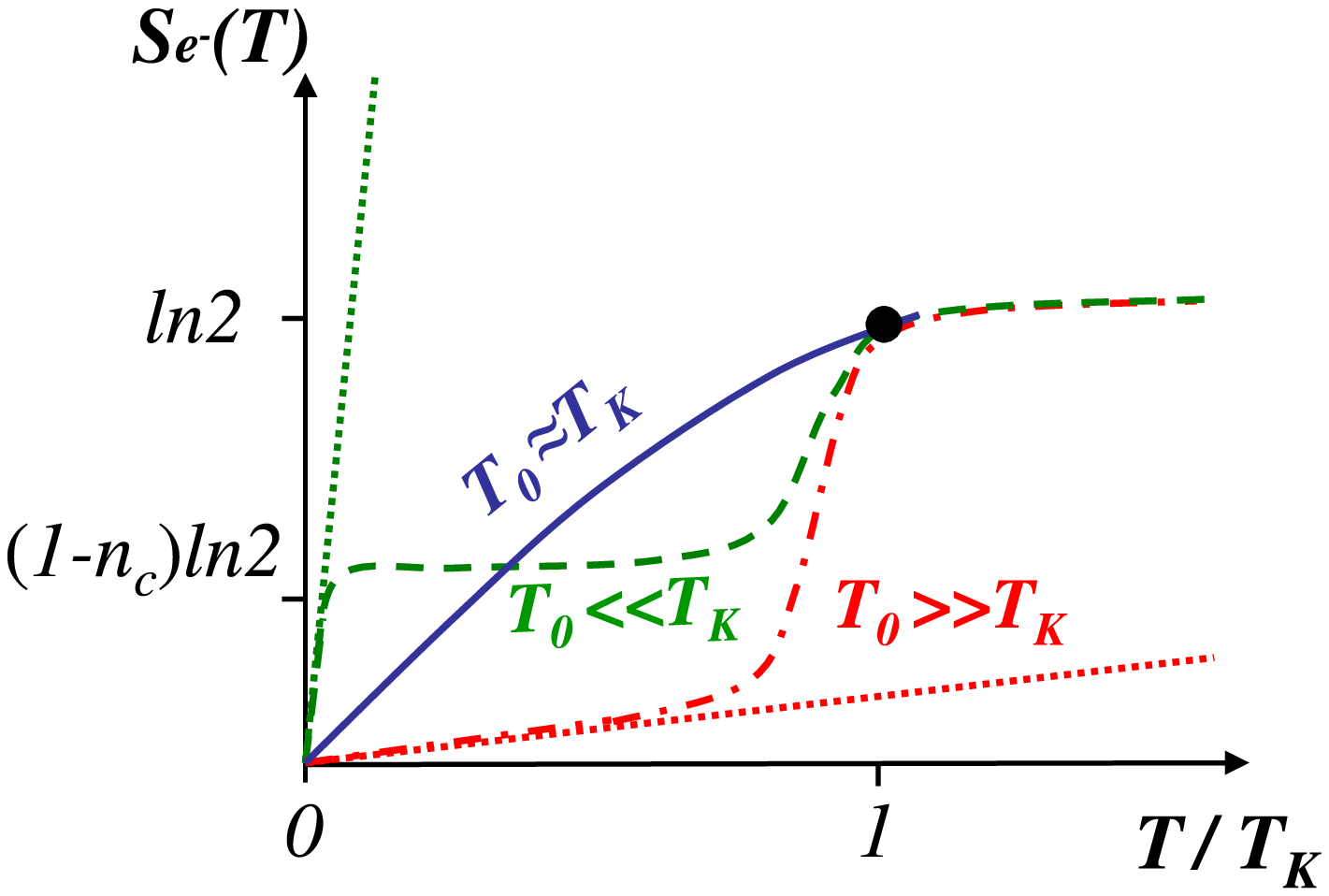}  
\vspace*{1cm}
\caption{(Color online) 
Schematic plot of the electronic contribution to  
the entropy $S_{e^{-}}$ as a function of the normalized temperature  
T/T$_{K}$ for three cases: 
T$_{0}\gg$T$_{K}$ (red dash doted line), 
T$_{0}\approx$T$_{K}$ (blue solid line), 
and $T_{0}<<T_{K}$ (green dashed line). 
The doted lines indicate the linear Fermi liquid regime 
$S_{e^{-}}(T)=T/T_{0}$, with a rescaled slope T$_{K}$/T$_{0}$.  
For T$>$T$_{K}$ the three curves are identical, reflecting 
the linear contribution from the conduction band 
$S_{e^{-}}(T)=\ln{2}+T/D$. 
The black dot refers to a standard determination of T$_{K}$ 
from the electronic entropy: $S_{e^{-}}(T_{K})=x\ln{2}$. 
On this schematic plot, we used $x=1$, which gives T$_{K}$ 
as defined by the slave boson ($r=0$) solution. 
Experimentally, where T$_{K}$ is defined by the crossover, 
one usually takes $x=1/2$.} 
\label{Schemasentropy} 
\end{figure} 
 
To find T$_K$ at finite $\rho_0(\mu)$ we use the Sommerfeld 
expansion of equation~(\ref{EqTK}) which gives in the 
$T_{K}<<D$ limit the result
\begin{eqnarray} 
T_{K}=\alpha_{0} (D^{2}-\mu_{0}^{2})^{1/2} 
F_{K} 
e^{-1/J(\mu_{0})\rho_{0}(\mu_{0})}~,  
\label{ExpressionTK} 
\end{eqnarray} 
where 
\begin{eqnarray} 
F_{K}
=
\exp{ 
\left[ { 
\int_{-(D+\mu_{0})}^{D-\mu_{0}} 
\frac{\rho_{0}(\mu_{0}+\omega)-\rho_{0}(\mu_{0})} 
{2|\omega |\rho_{0}(\mu_{0})}{d\omega}} 
\right] }~.    
\label{ExpressionFK} 
\end{eqnarray} 
$\omega$ is measured with respect to $\mu$, 
$\alpha_{0} =1.13$ is a numerical constant, and  
$\mu_{0}$ is defined by the integral  
\begin{eqnarray} 
\frac{n_{c}}{2} 
=\frac{n-1}{2} 
=\int_{-D}^{\mu_{0}}\rho_{0}(\omega)d\omega~. 
\label{Defmuzero} 
\end{eqnarray} 
By definition, 
$\mu_0$ is the Fermi level of $n_c$ non-interacting electrons 
which have a ``small''  FS. The result given by Eq.~(\ref{ExpressionTK})  
is derived in Ref.~\cite{burdin.2000} for a Kondo lattice  
with an analytic DOS. It also holds for $\rho_0(\omega)$ 
with an algebraic singularity  close to $\mu_{0}$.   
 
The Kondo temperature T$_{K}$, defined in Eq.~(\ref{EqTK}) characterizes  
a second order phase transition.  
We are aware that this transition is an artefact of the slave boson  
MF approximation and that an exact theory would lead to  
a crossover instead.  
Experimentally, the Kondo temperature T$_{K}$ is sometime estimated  
from the resistivity measurements, which show a maximum around T$_{K}$,  
or from the specific heat measurements, where T$_{K}$ is identified  
as the temperature at which the magnetic entropy (per impurity)  
becomes a substantial fraction of $\ln 2$, say  $S\simeq 0.5~\ln 2$ 
(see FIG.~\ref{Schemasentropy}). 
Note that with this definition of T$_{K}$ from the magnetic entropy 
we implicitly neglect the collective freezing of the entropy which
is related to the Ruderman-Kittel-Kasuya-Yosida interaction.

The fact that the Anderson lattice and the single impurity Anderson  
model have the same Kondo scale indicates that in both cases T$_K$  
separates the paramagnetic high-entropy phase from the low-entropy 
phase in which the conduction electrons start forming an incoherent 
screening clouds which reduce the local $f$ moment in each unit cell 
(see FIG.~\ref{FigSchemasDOS}). 
At temperatures at which the $c$ and $f$ states form a coherent band,  
and local screening clouds become correlated, the hybridization $V$ 
cannot be considered as a perturbation.  
For $T\ll T_K$,  the Hamiltonian has to be diagonalized by  
non-perturbative methods and the slave boson solution provides 
a reasonable description of the renormalized ground state. 
Close to the ground state, the low-energy excitations of  
a periodic system are Bloch waves and we expect them to be  
characterized by a FL scale T$_0$. 
The question is, how is T$_0$ related to T$_K$. 
 
\subsection{The Fermi liquid scale  $T_{0}$\label{T0}} 
\begin{figure}[h] 
\includegraphics[width=8.0cm,angle=90]{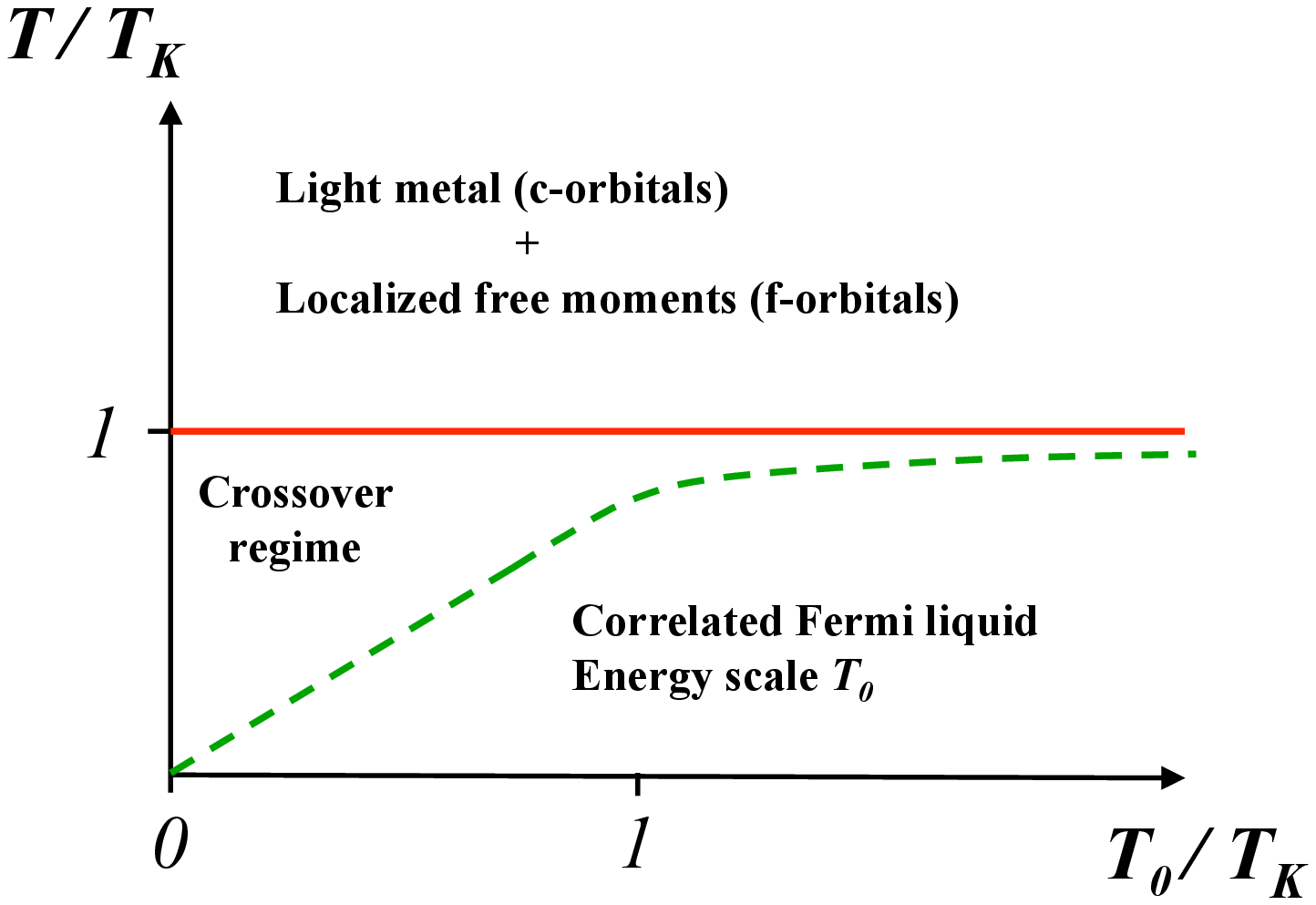} 
\vspace*{1cm}
\caption{(Color online) Schematic phase diagram of the PAM.  
The reduced temperature T/T$_K$ is plotted versus T$_{0}$/T$_{K}$,   
which is considered as a tunable parameter that can vary with  
the electronic filling, magnetic field and/or the shape of  
the non interacting DOS.  
The crossover regime can be non universal or non Fermi liquid. 
} 
\label{Schemasphasediagram} 
\end{figure}

The emergence of a strongly coupled $f-c$ fluid at T$_K$ 
does not imply that Kondo scale characterizes the behavior 
of such a fluid close to the ground state. 
Electrons described by Eq.~(\ref{PAMmagnetbosonsMF}) form  
at $T=0$ a coherent Fermi liquid with an energy scale $T_{0}$, 
which is defined in the absence of a magnetic field as, 
\begin{eqnarray} 
T_{0}\equiv \frac{1}{\rho^{\uparrow} (\mu)} 
=\frac{1}{\rho^{\downarrow} (\mu)}
~, 
\label{DefTzero} 
\end{eqnarray} 
where $\rho^{\sigma}(\mu)\equiv 
\rho_{c}^{\sigma}(\mu)+\rho_{f}^{\sigma}(\mu)$ is the 
renormalized density of electronic states at the Fermi level.    
The scale $T_{0}$ is relevant for the $T\rightarrow 0$ properties 
of the periodic Anderson model;  it determines 
the static spin susceptibility, $\chi_{loc}(T=0)\sim 1/T_{0}$, 
the specific heat coefficient $\gamma=C_V/T\sim 1/T_{0}$,  
and appears in the transport coefficients\cite{zlatic.08} 
which are given by simple powers of reduced temperature T/T$_0$. 
The slave boson result for T$_0$ is computed from 
Eqs.~(\ref{MFmagneteqr2} -- \ref{eq:rho_c}) at $T=0$, which gives 
\begin{eqnarray} 
\frac{E_{f}-\lambda-\mu}{2V^{2}} 
&=& 
\int_{-\infty}^{0} 
\frac{1}{\omega-\lambda} 
\rho_{0}\left({  
\omega+\mu-\frac{r^{2}V^{2}}{\omega-\lambda} 
}\right)  
d\omega~,  
\label{MFeqr3}\\ 
&&\nonumber\\ 
\frac{n_c+n_f}{2} 
&=& 
\int_{-\infty}^{0} 
\left[1+
\frac{r^{2}V^{2} }{(\omega-\lambda)^{2}}
\right]
\rho_{0}\left({\omega+\mu-\frac{r^{2}V^{2}}{\omega-\lambda}}\right)  
d\omega 
=
\int_{-\infty}^{\mu_L} 
\rho_{0}(\omega) d\omega 
~,\label{MFeqmu3} 
\end{eqnarray} 
where $\rho_{0}(\omega)$  is the unperturbed $c$ DOS and $\mu_{L}$ the chemical 
potential of $n=n_{c}+n_{f}$ non-interacting electrons. We have
\begin{eqnarray} 
\mu_{L} \equiv \mu+\frac{r^{2}V^{2}}{\lambda}
~  
\label{RelLuttinger} 
\end{eqnarray} 
which defines the shift in the chemical potential due 
to the enlargement of the Fermi volume of hybridized ($r\neq 0$) 
system with $n$ particles with respect to the Fermi volume of the 
non-interacting ($r=0$) band with $n-1$ particles, 
$\Delta\mu \equiv \mu_{L}-\mu = {r^{2}V^{2}}/{\lambda}~$.  
(This interpretation of $\Delta\mu$  neglects the width of the 
free-electron distribution function at temperature T$_K$ 
with respect to $D$ and assumes $\mu\simeq \mu_0$, which holds 
for $n_f\simeq 1$.) 
The low-temperature  FS is ``enlarged'' with respect to 
the high-temperature one, because it accommodates $n_f$ 
additional $f$ electrons. 
Equations ~(\ref{eq:rho_d}), (\ref{eq:rho_c}), and 
(\ref{RelLuttinger}) yield   
\begin{eqnarray} 
\rho(\mu) 
= 
\left[ 1+ \left( \frac{\Delta\mu}{rV} \right)^{2} \right] 
\rho_{0}(\mu_{L})~,  
\label{DOSTzero}  
\end{eqnarray} 
which shows that $\rho(\mu)$ and T$_0$ depend on the shape of the 
unperturbed $c$ DOS and the total number of particles. 
In the FL regime, the entropy behaves as $T/T_{0}$, 
the susceptibility is constant and transport coefficients 
are given by simple power laws of $T/T_{0}$. 
The schematic phase diagram of a system with $T_K\neq T_0$ 
is shown schematically in Fig.~\ref{Schemasphasediagram}. 

The effect of the $c$ DOS can be computed analytically in the 
$V<<D$ limit, assuming that $\mu$ is close to $\mu_0$.  
Since $\mu_{L}$  corresponds to $n$ non-interacting electrons 
("large" FS) and $\mu_0$ to $n_{c}\sim n-1$ electrons ("small" FS), 
we  approximate  $\Delta\mu\propto D$ (see FIG.~\ref{FigSchemasDOS}).  
In the limit $rV/D\ll 1$, the MF  
Eqs.~(\ref{MFeqr3}) 
and (\ref{MFeqmu3}) give   
\begin{eqnarray} 
\frac{E_{f}-\lambda-\mu}{2V^{2}} 
&=& 
{ \rho_{0}(\mu)\ln{ \left( { 
\frac{r^{2}V^{2}}{(D+\mu)\Delta\mu} 
}\right) } 
- 
\int_{-(D+\mu)}^{\Delta\mu} 
\frac{\rho_{0}(\mu+\omega)-\rho_{0}(\mu)}{ 
|\omega |}d\omega }
~,  
\label{MFeqr4}\\ 
&&\nonumber\\ 
\frac{n_c+n_f}{2} 
&=& 
\int_{-(D+\mu)}^{\Delta\mu}
\rho_{0}(\mu+\omega)d\omega~,  
\label{MFeqmu4} 
\end{eqnarray} 
where we neglected 
the corrections of order $(rV/D)^{2}$. 
Solving Eqs.~(\ref{MFeqr4}, \ref{MFeqmu4}) for $\lambda$ yields  
\begin{eqnarray} 
\lambda\simeq \frac{r^{2}V^{2}}{\Delta\mu} 
&\simeq& 
(D+\mu_{0}) 
~ F_{0} ~ 
e^{-1/J(\mu_{0})\rho_{0}(\mu_{0})}~,  
\label{Expressionlambda} 
\end{eqnarray} 
where 
\begin{eqnarray} 
F_{0}=\exp{ 
\left[ { 
\int_{-(D+\mu_{0})}^{\Delta\mu} 
\frac{\rho_{0}(\mu_{0}+\omega)-\rho_{0}(\mu_{0})}{ 
|\omega |\rho_{0}(\mu_{0})}{d\omega} 
}\right] }~.   
\label{ExpressionFzero} 
\end{eqnarray} 
The scale T$_0$ of a system of particles described by the periodic 
Anderson model close to the ground state is obtained from 
equations ~(\ref{DefTzero},\ref{DOSTzero},\ref{Expressionlambda}) as, 
\begin{eqnarray} 
T_{0}= 
\frac{r^{2}V^{2}}{(\Delta\mu)^{2}}\frac{1}{\rho_{0}(\mu_{L})} 
= 
\frac{D+\mu_{0}}{\Delta\mu} 
\frac{F_{0}}{\rho_{0}(\mu_{L})} 
e^{-1/J(\mu_{0})\rho_{0}(\mu_{0})}~.  
\label{ExpressionTzero} 
\end{eqnarray} 
The scales T$_0$ and  T$_K$ have the same exponential dependence 
on the coupling  constant $J(\mu_{0})$ but their pre-factors 
are not affected by $\rho_0(\omega)$ in the same way and can differ 
considerably. 
The ratio T$_0/$T$_K$ which is constant for a given set of parameters 
can be changed by applying pressure or magnetic field. 

At temperatures low with respect to 
$T_{FL}={\min} \left\{ T_{0},T_{K}\right\}$ the system behaves 
as a Fermi liquid (provided we are not too close to the 
Kondo insulating state) but  the crossover from the high-temperature 
to the low-temperature regime proceeds differently for T$_0 \ll$T$_K$  
than for T$_0 \gg$T$_K$. This is shown schematically 
in Figs.~\ref{Schemasentropy} and ~\ref{Schemasphasediagram}. 
If T$_{0}\ll$T$_{K}$, we have $T_{FL}=T_{0}$ and, for $T_{0}<T<T_{K}$, 
the system exhibits an extended non-universal behavior. 
If the electronic occupation $n_{c}$ is neither too small nor 
too close to half-filling, the MF solution of the slave boson 
equations~\cite{burdin.2000} predicts the entropy with a plateau 
at about $(1-n_{c})\ln{2}$ (see FIG.~\ref{Schemasentropy}),  
which characterizes $1-n_{c}$ unscreened magnetic ions.
If $T_{0}\gg T_{K}$, the high-temperature perturbative regime persists 
all the way down to T$_K$, where the properties change abruptly 
and the system enters the  FL phase characterized by T$_0$.
Only for $T_{0}\sim T_{K}$, the lattice system is characterized 
by a single energy scale, as in the single impurity case. 
 
 \subsection{Comparison of the Fermi liquid and the Kondo scales
\label{T0/TK}   } 
The Kondo coupling $J(\mu)$ defined by Eq.~(\ref{DefKondocoupling})  
is always smaller than the half bandwidth $D$ of the non-interacting  
$c$ DOS, such that T$_{K}$ and $T_{0}$ are exponentially   
small due to the factor  $~\exp\{-1/J(\mu_{0})\rho_{0}(\mu_{0})\}~$. 
The analytic expressions given by 
Eqs.~(\ref{ExpressionTK}, \ref{ExpressionTzero}) 
yield  
the result
\begin{eqnarray} 
\frac{T_{0}}{T_{K}} 
= 
\left( \frac{D+\mu_{0}}{D-\mu_{0}}\right)^{1/2} 
\frac{1}{\alpha_{0} \rho_{0}(\mu_{L})\Delta\mu} 
\frac{F_{0}}{F_{K}}~, 
\label{ExpressionTzerosurTK} 
\end{eqnarray} 
which does not depend on the Kondo coupling $J(\mu)$ but,  
as discussed previously~\cite{burdin.2000}, varies with the electronic 
occupation and with the shape of the non-interacting $c$ DOS.  
Electronic filling effects have been discussed using the MF analysis 
of the Kondo lattice~\cite{burdin.2000} 
and the DMFT solution of the periodic Anderson model\cite{PhysRevB.55.R3332}.  
In the limit $n_{c}\to 0$ ($n\approx 1$), the first factor  
$D+\mu_{0}$ vanishes and $T_{0}<<T_{K}$  
(see FIG.~\ref{FigSchemasDOS} (a)).  
Another physically relevant case is found in the limit $n_{c}\approx 1$,  
which corresponds to a Kondo insulator with the vanishingly small   
$\rho_{0}(\mu_{L})$. Equation ~(\ref{ExpressionTzerosurTK}) yields 
T$_{0}\gg$T$_{K}$ and the break-down of the FL laws is neither due to the 
proliferation of the quasi-particle excitations (controlled by $T_{0}$), 
nor to the thermal destruction of the Kondo screening (controlled by T$_{K}$),  
but is due to the proximity of the chemical potential to the 
hybridization gap in the DOS.  
 
We consider now in more detail the effects due to the shape-variation of  
$\rho_0(\omega)$.  
In order to separate this effect from the electronic occupation effects
~\cite{Nozieres.1985,Nozieres.1986},  
we assume that $n_{c}$ is close to $1$ (but not at half-filling exactly, 
so that the system is metallic).  
The first two factors on the r.h.s. of Eq.~(\ref{ExpressionTzerosurTK})  
are then of the order $1$, such that $T_{0}/T_{K}\approx F_{0}/F_{K}$.  
Assuming $\Delta\mu\approx D-\mu_{0}$ we obtain from Eqs.~(\ref{ExpressionFK}) 
and (\ref{ExpressionFzero}) a simple relation 
\begin{eqnarray} 
\frac{T_{0}}{T_{K}} 
\sim  
\exp{ 
\left[ { 
\int_{-(D+\mu_{0})}^{D-\mu_{0}} 
\frac{\rho_{0}(\mu_{0}+\omega)-\rho_{0}(\mu_{0})}{ 
2|\omega |\rho_{0}(\mu_{0})}{d\omega} 
}\right] }~,
\label{ApproxTzerosurTK} 
\end{eqnarray} 
which describes the dependence of $T_{0}/T_{K}$ on the specific  
form  of $\rho_0$. 
A constant $\rho_0$ gives $T_{0}\sim T_{K}$, which explains  
the T/T$_K$ scaling observed in some heavy fermion compounds. 
If $\mu_{0}$ is close to a local maximum of $\rho_0(\omega)$  
the integrand in Eq.~(\ref{ApproxTzerosurTK}) is negative  
in the main part of the integration range, such that $T_{0}\leq T_K$, 
as found in the systems with the 
`protracted screening'~\cite{PhysRevB.55.R3332,bauer:125102}.  
A sharp spike in $\rho_0(\omega)$ close to $\mu_0$ 
would exponentially reduce T$_0$ with respect to T$_K$. 
On the other hand, if $\mu_{0}$ is close to a local minimum  
[see FIG.~\ref{FigSchemasDOS} (b)] one finds $T_{0} \gg T_{K}$,   
which could be understood by the following intuitive argument.  
The incoherent Kondo cloud which begins to form at $T\approx T_K$  
involves a few conduction states around the Fermi energy $\mu_{0}$.  
These states are part of the conduction band with a "small" FS  
(the FS of $n_{c}=n-1$ non-interacting electrons).  
When temperature decreases much below T$_{K}$, the local $f$ 
orbitals hybridize  with the conduction electrons to form 
a coherent Fermi liquid which is characterized by a "large" Fermi 
surface (the FS of $n$ non-interacting electrons). 
Thus, $T_{0}$ is affected by all the states between the ``small'' 
and the ``large'' FS, as well as some additional holes inside  
the ``small'' FS~\cite{burdin.2000}.  
For $n_{c}\simeq 1$ and $\mu_{0}$ close to the minimum of $\rho_0(\omega)$,  
the low-temperature increase of the Fermi volume leads to the DOS 
which is much larger than the one used to evaluate T$_{K}$  
(see FIG.~\ref{FigSchemasDOS} (c)). In that case, the formation 
of the Fermi liquid is  "self-amplified", yielding $T_{0}\gg T_{K}$.  
We recall that for $T_0\neq T_K$ the FL regime sets in at 
temperatures that are the smaller than either T$_0$ or T$_K$. 

We are aware that corrections to MF slave boson analysis might 
occur from a more accurate treatment of the model; nevertheless, 
since the SB approximation is known to be correct at low energy, we 
expect such corrections to provide a similar integral expression, 
where the $1/\vert \omega\vert$ divergency would be smoothed out at 
hight frequencies. The analytical expression~(\ref{ApproxTzerosurTK}) 
still provides a good quantitative estimation of the band shape effect 
in the vicinity of the chemical potential. 
The relative magnitude of $T_{0}$ and T$_K$  
is related to the functional form of $\rho_0(\omega)$ around 
$\omega=\mu$.   
 
\subsection{{Effect of a magnetic field}\label{B}   }   
We next consider the slave boson solution in the presence of an 
external magnetic field which couples to  the $f$ orbitals. 
The direct effect on the $c$ electrons is neglected, even 
though they can be polarized due to the interaction with $f$ electrons.  
In the linear response regime, the local magnetization of the $f$ 
orbitals, 
$m_{z}(h)\equiv \frac{1}{2{\cal N}}
\sum_{i}(\langle f_{i\uparrow}^{\dagger}f_{i\uparrow}\rangle
-\langle f_{i\downarrow}^{\dagger}f_{i\downarrow}\rangle )$, 
is proportional to the applied field $h$ and, in the Kubo 
formalism~\cite{mahan.81}, the proportionality factor is equal to 
the local static susceptibility,  computed  in the absence of the
field. At $T=0$, the magnetization is $m_{z}(h)\propto h/T_{0}$, 
where $T_{0}$  is defined by Eq.~(\ref{DefTzero}) 
\begin{figure}[h] 
\includegraphics[width=7.0cm,angle=-90]{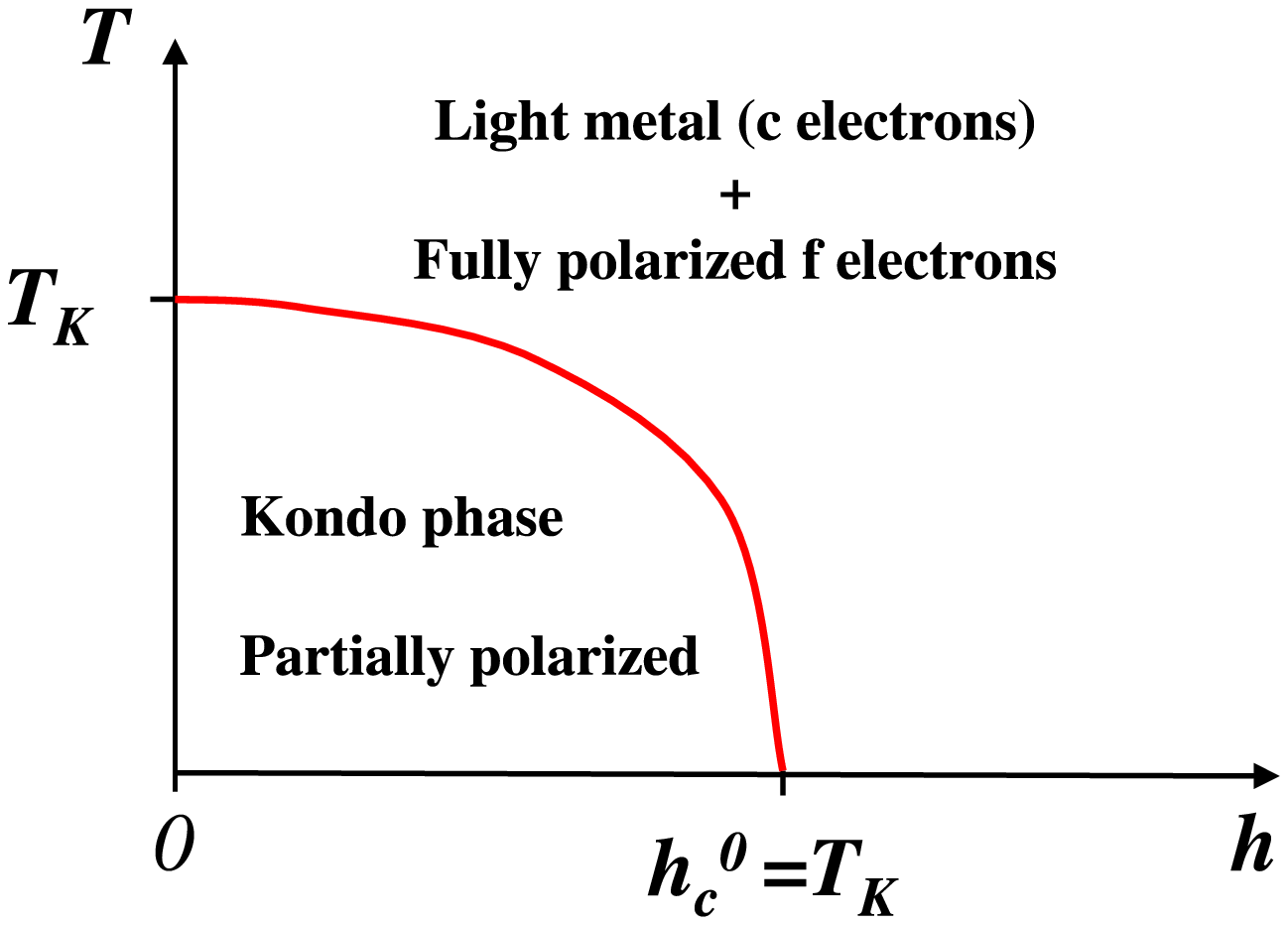} 
\vspace*{1cm}
\caption{(Color online) 
Schematic phase diagram of the PAM as a function of a magnetic field 
$h$. The red solid line indicates $h_{c}(T)$ which separates the trivial 
$r=0$ solution from the non-trivial one, $r\neq 0$. }
\label{Magneticphasediag} 
\end{figure} 
Going beyond the linear response, the critical magnetic field 
$h_{c}(T)$ is defined at a given temperature by the transition 
between the $r\neq 0$ and $r= 0$ state.  
Solving the MF equations ~(\ref{MFmagneteqr2}) --
(\ref{MFmagneteqmu2}) in the $r\to 0$ limit 
and using the definition of the Kondo coupling $J(\mu )$ in  
Eq.(\ref{DefKondocoupling}), 
we find for $n_{f}=1$ 
\begin{eqnarray} 
\frac{1}{J(\mu)} 
= 
\int_{-\infty}^{+\infty} 
\frac{\omega}{\omega^{2}-h_{c}^{2}}\rho_{0}(\omega+\mu) 
\tanh\left[ \frac{\omega}{2T}\right]d\omega~, 
\label{equation_TK_H}
\end{eqnarray} 
which yields $h_{c}\neq 0$ for any regular $\rho_{0}(\omega)$.  
Unlike the zero-field result, Eq.~(\ref{DivergentCouplingT=0}), 
which is logarithmically singular and gives $J_c=0$ for any finite 
$\rho_{0}(\mu)$, equation~(\ref{equation_TK_H}) yields $J_c\neq 0$ 
for any finite field. 
For $h\geq h_{c}(T)$, there is only the trivial $r=0$ solution 
which describes ferromagnetically polarized $f$-moments decoupled 
from the conduction band (see FIG.~\ref{Magneticphasediag}).  
For $h\leq h_{c}(T)$, the equations have a non-trivial $r\neq 0$ 
solution which describes a partially polarized Fermi liquid. 
At $T=0$, the weak coupling limit ($J\ll D$) yields the universal relation  
\begin{eqnarray} 
h_{c}^{0}\equiv h_{c}(T=0)=\alpha_{0}^{-1}
F_K  e^{-{1}/{J(\mu)\rho_0(\mu)}}
=\alpha_{0}^{-1} T_K~. 
\end{eqnarray} 
Since $\alpha_{0}=1.13$, we have $h_{c}^{0}\simeq T_{K}$. 
At finite temperatures, the solution of Eq.~(\ref{equation_TK_H})  
with constant $c$ DOS yields the critical line 
\begin{eqnarray} 
[h_c(T)/h_{c}^{0}]^2+[T/T_K]^2=1~,
\label{hc_boundary}
\end{eqnarray}   
which separates the trivial solution (the decoupled phase) from the 
non-trivial one (the FL phase) and holds for any  $J(\mu)$ and filling $n$.  
The critical line $h_c(T)$ obtained for a constant $c$ DOS is 
represented schematically in Fig.~\ref{Magneticphasediag}.  
The same relation is also found at half-filling, for any $c$ DOS. 
In general, we  expect a `nearly' universal phase boundary, 
with some small deviations due to  the structure of the  
$c$ DOS {\it and} the particle-hole asymmetry.  
\begin{figure}[h] 
\includegraphics[width=7.0cm,angle=-90]{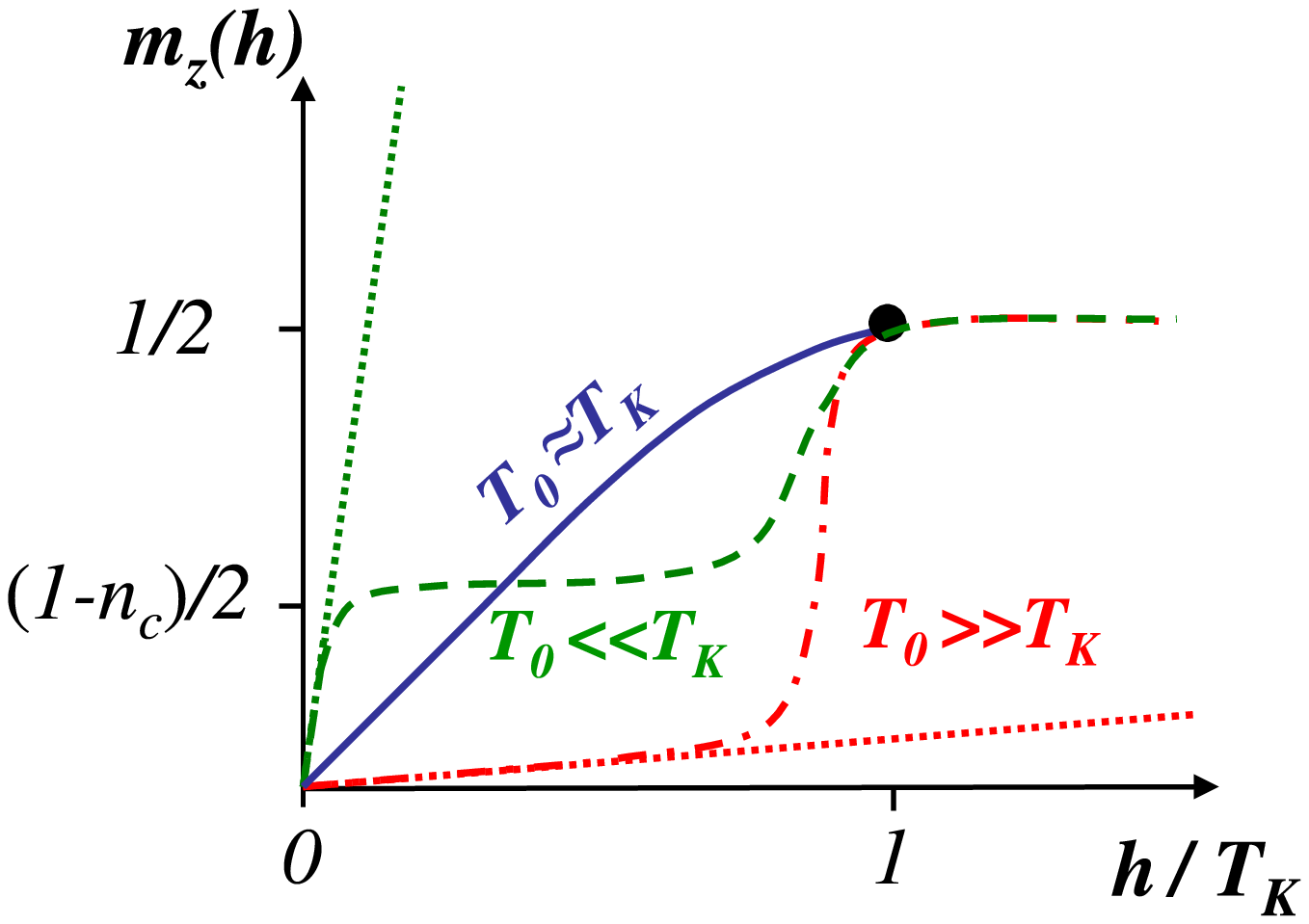} 
\vspace*{1cm}
\caption{(Color online) Schematic plot of the magnetization $m_{z}$ 
as a function of reduced magnetic field $h/T_{K}$ 
for T$_{0}\gg$T$_{K}$ (red dash doted line), 
T$_{0}\approx$T$_{K}$ (blue solid line), and 
T$_{0}\ll$T$_{K}$ (green dashed line). 
The doted lines indicate the  initial slope in the 
linear response regime, where $m_{z}(h)=h/T_{0}$. 
The black dot refers to the complete polarization 
of the local $f$ electrons, with $m_{z}=1/2$, which 
occurs a the critical field $h_{c}^{0}=T_{K}$. }
\label{SchemasMagnetization} 
\end{figure} 
%
Studying the development of the system as a function of temperature 
at constant field gives the critical temperature $T_K(h)$. 
Changing the field at constant temperature gives $h_c(T)$. 

The magnetization $m_{z}(h)$ 
obtained from the slave boson solution 
at $T=0$ is plotted in Fig.~\ref{SchemasMagnetization} as a 
function of reduced magnetic field $h/T_K$ for three typical cases: 
T$_{0}\ll$T$_{K}$, $T_{0}\approx T_{K}$, and $T_{0}\gg T_{K}$. 
A non-constant $c$ DOS leads to different types of magnetization 
curves which resemble the temperature dependence of the 
entropy depicted in Fig.~\ref{Schemasentropy} 
(with the T/T$_K$ axis replaced by $h/h_{c}^{0}$).  
For $h< \min~\{ T_{0}, T_{K}\}$, the system is in the linear response 
regime, shown in FIG.~\ref{SchemasMagnetization} by dotted lines; 
the slope of  $m_z(h)$ is T$_{K}$/T$_{0}$, when plotted 
versus $h/T_K$.  
This regime is analogous to the FL regime described from the entropy. 
In higher fields,  the behavior of the slave boson solution depends 
on the ratio $T_{0}/T_{K}$.  
For T$_{0}\ll$T$_{K}$, the $T=0$ magnetization is linear for small 
fields and then saturates.  For $T_{0}< h< T_{K}$, 
the magnetization might have a plateau which signifies a saturated FL 
($r\neq 0$) with a 'large' FS. At very high fields, 
$h \geq  T_{K}\sim h_{c}^{0}$, the magnetization becomes a universal function 
of $h/h_{c}^{0}$. 
In the opposite case, $T_{K}\ll T_{0}$, the low field limit gives  
$m_{z}(h)\propto h/T_{0}$ and the initial slope appears very small 
when plotted versus $h/h_{c}^{0}$.  
Once the field exceeds the critical value, $h_{c}^{0}\sim T_{K}$, 
the local moments are unscreened and $m_{z}(h)$ rises rapidly towards 
the free-ion value.  Thus, at about $h\simeq h_{c}^{0}$ such systems 
exhibit a meta-magnetic transition from an unpolarized Fermi liquid 
to the polarized spin-lattice (the r=0 phase).  
Of course, this simple considerations should be corrected for  
direct and indirect effects due to the conducting sea.  
 
\subsection{Transport coefficients in the FL regime\label{transport} }  

The slave boson Hamiltonian in Eq.~(\ref{PAMmagnetbosonsMF}) 
provides the approximate FL scale T$_0$ and the low-temperature 
thermodynamics but has no relaxation mechanisms that could lead to 
stationary heat and charge currents. To calculate the 
transport properties of the $SU(N)$ symmetric Anderson model 
in the  $T\to 0$ limit we use the Fermi liquid theory\cite{yamada.86} 
which takes into account the quasiparticle (QP) damping and leads 
to  a finite relaxation time. 
The QP excitations of the full model have the same dispersion 
as the excitations of the MF slave boson model but are restricted 
to the immediate vicinity of the Fermi level. 
In the $\omega\to 0$  limit,  where the imaginary part of the 
$f$ electron self energy $~\Sigma_f(\omega)$ can be neglected, 
the QP and the slave boson dispersion assume the same form. 
The correspondence is obtained by identifying $rV$ with 
$~\sqrt{Z_f}$ and $\lambda$ with $~\tilde\omega_f$, where 
$~Z_f^{-1}=[1-\partial\Sigma_f/\partial\omega|_{\omega=0}]$ 
is the renormalization factor,  
$~\tilde\omega_f = [E_f+{\rm Re}\Sigma_f(0)-\mu]Z_f$ 
the renormalized position of the $f$ level, 
and the excitation energies $\omega$ are measured in both cases 
with respect to the renormalized chemical potential $\mu$. 
Unlike the infinitely long-lived MF excitations, which are formally 
defined for $\omega\leq D-\mu$, the QP excitations are defined for 
$~{\rm Im }~\Sigma_f(\omega) \ll \omega^2$. 

We calculate the transport coefficients of the periodic Anderson 
model with constant hybridization using the fact that the charge and 
energy current density  operators 
satisfy the Jonson-Mahan theorem\cite{mahan.98}.  
This allows us to express the charge conductivity by  
$\sigma(T)= e^2 N L_{11}$, the thermopower by  
$\alpha(T) |e| T = -  L_{12}/ L_{11}$, 
and the electronic contribution to the thermal conductivity by 
$\kappa(T) T = N ( L_{22}   -  L_{12}^2/ L_{11} )$. 
In each of these expressions we have introduced 
the transport integrals: 
\begin{equation} 
\label{mj} 
L_{mn} 
= 
\int d\omega 
\left(-\frac{dn_{F}}{d\omega}\right) 
{\omega}^{m+n-2} \Lambda(\omega,T) . 
\end{equation} 
where $n_{F}(\omega)=1/[1+\exp(\beta\omega)]$ is the Fermi-Dirac 
distribution function and $\Lambda(\omega,T)$ is defined by 
the Kubo linear response theory\cite{mahan.81}. 
At low temperature $(-dn_{F}/d\omega)$ approaches delta function  
and the main contribution to the integrals in Eq.~(\ref{mj})  
comes from the low-energy excitations within the Fermi window,  
$|\omega|\lesssim T$. 
In the $\omega,T\rightarrow 0$ limit, a straightforward 
calculation yields for the three-dimensional systems\cite{zlatic.08}  
\begin{equation} 
                             \label{lambda-FL} 
{\Lambda(\omega,T)} 
= 
\frac{1}{3}{v^2_F}  \rho_c(\omega)\tau(\omega,T),     
\end {equation} 
where we introduced the unrenormalized velocity 
$v_{\bf k} =- \nabla_{\bf k} \epsilon_{\bf k} $ 
and  denoted by ${v^2_F} $ the average of  $v_{\bf k}^2$ over  
the renormalized Fermi surface of hybridized states.  
The renormalized $c$ DOS is $\rho_c(\omega)$ and the relaxation 
time is $1/\tau(\omega,T)\simeq ~{\mathrm Im}~\Sigma_c(\omega)$, where 
$\Sigma_c(\omega)$ is the self energy of the {\it c} electrons 
which must include the quasiparticle damping.
The integrals in Eq.~(\ref{mj}) are evaluated by the Sommerfeld expansion 
(for details see Ref.~\cite{zlatic.08})  which yields the transport 
coefficients of the periodic Anderson model as simple powers of reduced 
temperature T/T$_0$. The pre-factors of various powers are functions 
of $\rho_0(\mu_L)$, $n_f$, and $\Delta\mu$ which depend on the local 
self energy $\Sigma_f(\omega)$ which is difficult to obtain for the 
excitation energies within the Fermi window, $|\omega|\leq T$.  
To avoid this problems, we replace T$_0$ and all other renormalized 
quantities that appear in the FL expressions by the slave boson MF results. 

The FL result for the electrical resistivity can be written 
as~\cite{zlatic.08}
\begin{eqnarray} 
R(T) 
&\simeq& 
                             \label{eq: rho_gama} 
\frac { 9^2 (\Delta\mu/n_f)^2 } { N(N-1)\pi e^2 {v^2_F} } (\gamma T)^2  
=
\frac { 9 \pi^3 (\Delta\mu/n_f)^2 ~T_K^2}{ N(N-1) e^2 {v^2_F}~T_0^2} 
\left(\frac {T}{T_K}\right)^2
=
A~T^2 ~,
\end {eqnarray} 
where $\gamma=(\pi^2/6)N\rho(\mu)=\pi^2/3 T_0$  and  
the in the second equality $R(T)$ is expressed on the 
reduced temperature scale T/T$_K$. The coefficient $A=R(T)/T^2$  
depends not only on specific heat coefficient $\gamma$ but 
on the difference in the chemical potentials of unhybridized 
and hybridized Bloch states, the effective degeneracy of the model, 
and the square of the Fermi velocity averaged over the hybridized FS. 
This result can be used to explain the pressure- and magnetic 
field-dependence of $A$, which has been studied experimentally 
in various systems.

The FL result for the Seebeck coefficient is given 
for $n_f\simeq 1$ by the expression 
\begin{equation} 
                             \label{alpha-tau} 
{\alpha(T)} 
= 
\mp 
\frac{12~\gamma ~T}{n_f|e|}  
=
\mp 
\frac{4\pi^2 ~ T_K}{n_f|e|~T_0}  
\frac{T}{T_K} ~.
\end {equation} 
Since the doubly occupied {\it f} states are removed from  
the Hilbert space, the model is highly asymmetric, and the  
initial slope $\lim_{T\to 0} \alpha(T)/T$ never vanishes.  
In bad metals with a low-carrier concentration,  
$\alpha(T)$ could be very large.  

The thermal conductivity in the FL regime reads, 
\begin{equation} 
                                       \label{kappa2} 
\kappa(T) 
= 
T \sigma(T) 
{\cal L}_0(T) 
\end {equation} 
which yields in the $T\to 0$ limit the Wiedemann-Franz (WF) law,  
$\kappa(T)\propto \ T \sigma(T)$.  
However, the usual Lorentz number, $ {L}_0={\pi^2 }/{3e^2}$,   
is here replaced by the effective one,  
\begin{equation} 
                                       \label{Lorentz} 
{\cal L}_0(T)=\frac{\pi^2 }{2e^2} 
\left[1- \frac{32\pi^2}{ n_f^2} \left( \frac{T}{ T_0}\right)^2\right] ~.  
\end {equation}  
The correction given by the square bracket could lead to 
substantial deviations from the WF law much below T$_0$, 
since the factor multiplying the $T^2$ term is very large.

\section{Discussion of the experimental data \label{results} } 
 
The slave boson solution of the periodic Anderson model 
is used in this Section to discuss the effects of the band structure 
on the properties of the intermetallic compounds with 4$f$ ions. 
First, we consider the case of the $c$ DOS with a maximum 
in the vicinity of the chemical potential, such that the Fermi 
liquid scale is much smaller than the Kondo scale, T$_0\ll$T$_K$.  
These results provide a qualitative explanation of the experiments 
on YbAl$_3$ or YbMgAl$_4$.
The opposite case, T$_0\gg$T$_K$, occurs when the chemical potential 
is close to the minimum of the $c$ DOS;  these results explain 
the main experimental features of YbInCu$_4$-like compounds. 
Finally, for T$_0\simeq$T$_K$, which seems to characterize 
CeCu$_2$Si$_2$ and CeCu$_2$Ge$_2$, we discuss the rapid 
variation of the $T^2$ coefficient of the electrical 
resistance and of the residual resistance with pressure. 
The anomalies are related to the pressure-induced change 
of the effective degeneracy of the $f$ states. 

\subsubsection{The  $T_0 \ll T_K $ case \label{T0<<TK} } 

Unlike the experiments on dilute alloys,  
the overall temperature dependence of the experimental data on 
YbAl$_3$, Yb$_{1-x}$ Lu$_x$ Al$_3$~\cite{bauer:125102} for $x\leq 0.05$, 
YbMgCu$_4$ and YbCdCu$_4$\cite{lawrence.PhysRevB.63.054427}, 
cannot be explained by a single energy scale. 
In these compounds, the high-temperature resistivity is a large 
and slowly varying function of temperature\cite{vandaal.74,rowe.2002},  
the thermopower\cite{vandaal.74,rowe.2002} has a broad peak around 350 K, 
the magnetic susceptibility and the specific heat have a broad 
maximum at $T_{max}\simeq 125$ K, and the Hall coefficient is typical 
of a metal in which the $c$ electrons scatter on local moments
~\cite{Cornelius.2002,bauer:125102}. 
A similar high-temperature behavior is seen in YbXCu$_4$ 
(X=Cd, Mg, Zn)~\cite{lawrence.2000}. 
The inelastic neutron scattering data on YbAl$_3$ show 
a broad Lorentzian spectrum centered at about 
540 meV~\cite{Murani.1985,Osborn.1999}, which can be understood
in terms of the Kondo effect with T$_{K}\geq$ 500 K~\cite{Cornelius.2002}. 
Below 100 K, the electrical resistivity of YbAl$_3$ decreases 
due to the formation of a coherent ground state and 
for $T\leq$T$_0\simeq$ 40 K the resistivity is a parabolic and 
the thermopower a linear function of temperature 
~\cite{vandaal.74,rowe.2002,Ebihara.2000}, indicating a Fermi liquid.
For $T\leq T_{0}$, the zero-field anomalies in $\chi(T)$ and $C_V(T)$ 
yield an enhanced effective mass~\cite{Cornelius.2002} 
of the order of 1/T$_0$ and the neutron scattering 
data~\cite{Murani.1985,Osborn.1999} show a narrow peak 
at about 30 meV, which is due to a hybridization gap. 
However, if we plot the low-temperature transport coefficients 
on a reduced temperature scale T/T$_K$ they appear to be strongly 
enhanced with respect to the predictions of the single impurity 
Anderson model with Kondo scale $T_K\simeq $ 500 K.  
Neither the $A$ coefficient of the resistivity nor the slope of 
the thermopower can be explained by the single impurity c
calculations that would capture the main features of the 
high-temperature data and give $T_K\simeq $ 500 K. 
Optical conductivity at 7 K shows a narrow Drude-like response 
that is often found in heavy fermion systems
\cite{Okamura.2003,okamura:jpsj_76} and another mid-infrared peak (MIR) 
that can be associated with the hybridization gap. 
The optical spectra do not change appreciably for 
7 K $\leq$T$\leq$ 40 K, as expected of a system with the 
characteristic energy scale  T$_{0}\simeq 50$ K. 
However, the Drude peak broadens and the MIR peak vanishes 
at higher temperatures.  
The de Haas - van Alphen experiments performed up to $h\simeq 40$ T, 
show that the effective mass is reduced along certain directions 
in {\bf k}-space by a factor of 2 without a significant alteration 
of the shape of the Fermi surface~\cite{Ebihara.2003}.  
The fact that the low-temperature susceptibility anomaly 
is suppressed in the fields of about 40 T indicates that 
the high-field mass renormalization and the zero-field 
anomalies below $T_{0}$ describe different aspects of 
the same physical state. 
The low-temperature anomalies are easily destroyed by disorder
(they vanish in Yb$_{1-x}$ Lu$_x$ Al$_3$ for $x\geq 0.05$),  
which is another indication that they are related to the 
coherent state~\cite{Ebihara.2003}. 

The overall shape of the experimental data and different 
characteristic energy scales found at high and low temperatures 
can be explained by the slave boson solution of the 
periodic Anderson model close to half-filling. 
As shown in Secs.~\ref{T0/TK} and \ref{B}, if the chemical potential 
is close to the maximum of the unperturbed $c$ DOS, the mean field equations 
yield $T_0\ll T_K  \simeq h_{c}^{0} $ and give rise to 
a 'slow crossover' from the LM to the FL phase.    
Taking $T_0\simeq 50$ K and $T_K\simeq 500$, 
as suggested by the experiment, we find 
that the low-temperature susceptibility and 
the specific heat coefficient are FL-like 
and much enhanced with respect to the expectations 
based on the behavior in the incoherent regime. 
The FL results for the $T^2$ term of the resistivity 
(see Sec.~\ref{transport}) explain the enhancement that 
one finds below 50 K when $\rho(T)$ is plotted on the T/T$_K$ scale.  
The low-temperature thermopower 
${\alpha(T)} \propto\left({T_K}/{T_0}\right) {T}/{T_K}$
is also enhanced by T$_K/$T$_0$ with respect to the predictions 
of the single impurity calculation that reproduce the 
thermopower maximum above 300 K. 
For temperatures between T$_0$ and T$_K$, the magnetization 
and the magnetic entropy are reduced with respect to the 
free-ion value, which is only recovered for $T > T_K$. 
The slave boson solution gives $m_{z}(h)\simeq  h/T_0$ for $h\ll T_0$ 
and $m_{z}(h)\propto$ const for $h\gg T_0$. 
Since $T_0\ll h_{c}^{0}$, the slave boson order parameter 
is finite for $h\geq T_0$, i.e., for $T_0\leq h \leq T_K$, 
the system is a polarized heavy FL with a `large FS'. 
Such a behavior explains de Haas -- van Alphen data which show 
that the FS does not change much up to the fields of about 40 T. 

The model with the chemical potential close to the maximum in the 
$c$ DOS describes Yb$_{1-x}$ Lu$_x$Al$_3$ for $x\leq 0.5$.  
For higher concentrations of Lu there are so many additional holes 
in the conduction band that the chemical potential shifts away 
from the peak in the $c$ DOS. In that case, T$_0$ and T$_K$ approach 
each other and the 'slow' crossover does not occur. 
For $x>0.5$, one can expect a crossover from a lattice regime 
with T$_{0}\neq$T$_{K}$ to a universal dilute regime 
with T$_{K}=$T$_{0}$, as discussed in Ref\cite{burdin-fulde.2007}. 

The slave boson dispersion defined by the Hamiltonian in  
Eq.~(\ref{PAMmagnetbosonsMF}) explains the Drude and the MIR peaks 
found at 7 K in the optical conductivity. 
However, we hesitate to discuss the temperature dependence 
of the hybridization gap using the mean field results.   
These results are obtained by enforcing the constraint $n_f=1$ 
only on the average, so that the auxiliary fermions 
are mapped on a free electron gas. 
The proper redistribution of the spectral weight should not 
neglect the QP damping and should use the solution which is 
valid at all energy  scales. 

\subsubsection{The  T$_0 \gg$T$_K$ case \label{T0>>TK} } 

Unlike YbAl$_3$ or Yb$_{1-x}$ Lu$_x$ Al$_3$ for $x\leq 0.5$,
the transition between the LM and the FL phase in
Yb$_{1-x}$ Lu$_x$ Al$_3$ for $x\geq 0.7$,
YbTlCu$_4$\cite{lawrence.PhysRevB.63.054427},
YbInCu$_4$ and Yb$_{1-x}$Y$_x$InCu$_4$ 
for $x\leq 0.5$~\cite{sarrao.1999} 
takes place at temperature T$_v$ in an abrupt way.
In YbInCu$_4$, which we take as a typical example, 
a first order valence change (VF) transition occurs 
at ambient pressure at temperature T$_v$=40 K.
Above T$_v$,  the Yb ions are in the 3$^+$ configuration and
the magnetic response can be understood assuming an independent 
4$f$ hole in each unit cell. The ensuing high-temperature 
effective moment is then close to the free ion value, 
as observed experimentally.
The electrical resistance is very large and nearly temperature independent.
The Kondo scale deduced from these data is so much smaller than T$_v$,
that the $f$ and $c$ states are effectively decoupled for T$\geq$T$_v$.
The magnetic fields up to 40 Tesla do not produce any appreciable
magnetoresistance which is also an indication of a small Kondo coupling.
The large values of the electrical resistance and small Hall constant
cannot be explained in terms of the spin disorder scattering but can be 
taken as an evidence that the $c$ DOS has a pseudogap or a deep
minimum in the immediate vicinity of the chemical potential.
When the compound is cooled down to temperature T$_v$, the lattice
expands and the Yb configuration changes from 3$^+$ to 2.9$^+$.
In the VF phase, the susceptibility and the specific heat 
coefficientare moderately enhanced ($\gamma\simeq 50$ mJ/mol K),
the resistivity is quadratic and the thermopower linear in temperature.
The ratio $\alpha/T\gamma$ is typical of a normal FL\cite{behnia.04} 
but the Kadowaki-Woods ratio is anomalously low.
The characteristic energy scale in the VF phase is much larger than T$_v$; 
the data give $T_0\approx 500$ K~\cite{dallera.2002}.
The optical conductivity\cite{garner.2000},
the Hall effect\cite{figueroa.1998}, and the thermoelectric
power\cite{ocko.2002} indicate that a bad metal with only a few 
states close to the chemical potential is transformed at T$_v$ 
into a good metal with small $\gamma$. That is, the VF transition 
is accompanied by a major reconstruction of the conduction states.

An application of pressure shifts $T_v(P)$ to lower
temperatures~\cite{mito.2003,mitsuda.2004,park.2006,mito.2007}
without changing the properties of the high-temperature state.
For $T\geq T_v(P)$, the susceptibility and the specific heat of
YbInCu$_4$ can be explained by the crystal field (CF) theory of 
independent $f$ states\cite{aviani.2008} split by 
$\Delta_{CF} \simeq $ 40 K into an excited quartet and 
two nearly degenerate doublets. This CF scheme agrees 
with the neutron scattering data\cite{severing.1990}. 
The entropy of the LM phase, obtained by integrating $C_V(T)/T$ 
\cite{park.2006,aviani.2008}, decreases from $S\simeq R\ln 8$ 
to $S\simeq R\ln 4$ as the system is cooled down to $T_v(P)$, 
as expected for two CF quartets without any Kondo screening.  
This indicates, once again, that the Kondo scale of the 
LM phase is much smaller than $T_v(P)$.
Pressure stabilizes the paramagnetic phase and at a 
critical pressure of $P_c=$ 2.5 GPa the LM persists 
down to $T_c=$ 2.5 K, where the magnetic transition 
removes the paramagnetic entropy of about $S\simeq R\ln 4$. 
The NMR~\cite{mito.2003,mito.2007} and neutron scattering data
show that the local moments in the magnetically ordered (MO) phase
are somewhat smaller than in the LM phase just above $T_N$. 
Since the specific heat coefficient and the $A$ coefficient 
of the resistivity are much larger for $P\geq P_c$ than for $P\leq P_c$, 
we speculate that the transition in the MO phase is accompanied 
by an increase of Kondo coupling.
As regards the doping, replacing Yb with Y or Lu
ions\cite{zhang.2002,mitsuda.2004} produces similar effects as pressure, 
i.e., doping stabilizes the LM configuration and shifts $T_c$ to lower 
temperatures.
In Eu-based intermetallics, the valence-change transition exhibits similar
features~\cite{wada.1997,mitsuda.2000}  as in YbInCu$_4$,  except T$_v$
is higher ($T_v\geq 100$ K)  and the  valence state of Eu ions changes 
almost completely [from $2^+$ ($f^7$) to $3^+$  ($f^6$)].

We close this experimental summary by mentioning that the VF 
transition shifts in an magnetic field to lower tempeatures. 
At ambient pressure, the critical field of $h_{c}^{0}\simeq 40$ 
Tesla suppresses the VF transition and removes completely 
the FL statate of YbInCu$_4$.
The experiments give $\mu_{eff} h_c^0 = T_v$, where $\mu_{eff}$ is the 
effective magnetization of the 4$f$ state in the LM regime, 
i.e., at the VF transition, the magnetic energy of the paramagnetic 
$f$ states is comparable to the Kondo energy of the FL phase. 
The critical field is temperature dependent and the phase boundary between 
the LM phase and the FL  phase is given by the expression 
$h_c(T)=h_c^0\sqrt{1-(T/T_v)^2}$, 
where $h_c^0$ is the zero-temperature value.

The aforementioned behavior of YbInCu$_4$ and similar systems 
can be understood from the slave boson solution of the periodic 
Anderson model. Taking the chemical potential of the 
high-temperature phase close to the pseudo-gap of 
the unrenormalized $c$ DOS we find a very smal Kondo 
temperature (see Sec.~\ref{TK}), which explains 
the Curie-like susceptibility, 
a large and nearly temperature-independent resistivity,  
small thermopower, and a negligible magnetoresistance 
observed for $T\geq T_v$. 
Since the Kondo screening is absent in the LM phase, 
we can neglect the hybridization altogether and discuss 
the high-temperature properties of YbInCu$_4$ by adding 
the Falicov-Kimball term to the effective Hamiltonian. 
This term opens a gap or a pseudogap in the excitation 
spectrum\cite{freericks_zlatic.1998} and explains most 
of the qualitative features in a self-consistent way.
For a quantitative description of the paramagnetic phase,  
one would have to include the corrections due to the 
CF splitting of the $f$ states\cite{freericks_zlatic.2003}.  

The peculiar feature of YbInCu$_4$ and similar systems 
is that the local moments remain unscreened 
down to very low temperatures. System with a pseudogap 
in the $c$ DOS cannot remove the paramagnetic entropy by 
Kondo effect but have to approach the ground state by 
an entirely different route.
In YbInCu$_4$ the transition into a low-etropy state is 
achieved by an iso-structural phase transition which expands 
the lattice and facilitates the valence fluctuations between 
xthe low-volume 4$f^{13}$ and the large-volume 4$f^{14}$ 
state of Yb.
The presence of the 4$f^{14}$ configuration in the ground state 
means that some $f$ holes are transferred in the $c$ band, 
so that the chemical potential is shifted away from the 
pseudogap. 
This increases the Kondo coupling and makes the Kondo 
temperature of the expanded lattice comparable to T$_v$.  
For $T\leq T_K$, the local moment disappears, because 
the $f$ and $c$ states form hybridized bands, 
i.e.,  the $f$ holes participate in the Fermi volume.
The $f$--$c$ charge transfer induced by hybridization stabilizes
the low-entropy FL state and compensates the loss of elastic 
energy due to the lattice expansion. Once the Kondo scale of 
the hybridized system is equal to T$_v$, the expansion terminates, 
as can be seen from the following argument. 
The VF transition takes place at T$_v$, where the free
energy of the high-temperature phase, ${\cal F}_{LM}$, is equal to
the free energy of the low-temperature (expanded) phase, ${\cal F}_{FL}$.
Below T$_v$, the free energy is dominated by the hybridization 
term in the internal energy which stabilizes the FL state, 
${\cal F}_{FL} < {\cal F}_{LM}$.  
Above T$_v$, where the free energy is dominated by the paramagnetic entropy, 
the local moments of the pseudogapped phase are completely free,
while in the expanded lattice they would be partially screened. 
Since the free energy of the expanded lattice would be bigger 
than ${\cal F}_{LM}$, the most stable configuration above T$_v$ 
is the low-volume (paramagnetic) one. 

In the expanded lattice, the chemical potential is still rather 
close to the minimum of the $c$ DOS, which makes the FL scale 
much larger than the Kondo scale T$_v$ (see Sec.~\ref{T0}).
If we estimate the FL corrections to the $T=0$ value of the magnetic
susceptibility or the electric resistivity up to ${\cal O}[(T/T_0)^2]$
and assume $T_0/T_K\leq 10$ (as indicated by the data),
the maximum relative deviation at $T\sim T_v$ is  $1$\%.
The Kadowaki-Woods ratio in the FL phase is anomalously 
small, because the CF splitting does not affect the delocalized 
$f$ states, so that the $f$ states are effectively 8-fold degenerate. 
The FL analysis presented in Sec. III.3, shows that the $A$ 
coefficient is then reduced by a factor $1/N(N-1)$. Thus, the 
slave boson theory provides an overall description of YbInCu$_4$-like 
compounds at ambient pressure.

To account for the effects of pressure, we recall that 
pressure stabilizes the low-volume (4$f^{13}$) configuration 
with respect to the large-volume (4$f^{14}$) one and model this effect 
by shifting the $f$ level away from the chemical potential.
This reduces the $f$--$c$ coupling and the Kondo scale,  
shifts T$_v$ to lower temperatures and removes eventually 
the VF transition. 
For very large pressures, the paramagnetic entropy of 
the LM phase is not removed by the the lattice expansion 
but by a transition into a magnetically ordered (MO) phase 
which takes place at Curie temperature $T_c$. 

In the presence of an external magnetic field, the condition for the phase
boundary,  ${\cal F}_{LM}(T,h)={\cal F}_{FL}(T,h)$, can be approximated
by the condition $S_{LM}(T_v,h)=$ const., where $S_{LM}(T_v,h)$
is the entropy of the LM phase\cite{dzero.2000}.
This approximation uses the fact that T$_0$ is not affected 
by the pseudo-gap, so that $T_0\gg T_v$, and the temperature 
and the field-dependence of ${\cal F}_{FL}$ can be neglected 
for $T\leq$T$_v(h)$. 
The critical line obtained in such a way agrees very well 
with the experimental data\cite{dzero.2002b,freericks_zlatic.2003} 
and with the slave boson result given by Eq.~(\ref{hc_boundary}).
For large fields, the slave boson solution shown by the 
dashed-doted curve in Fig.\ref{SchemasMagnetization} 
explains the metamagnetic transition that takes place
at the critical field, $h_{c}^{0}\propto T_v$.


Similar reasoning explains also the unusual behavior of
the newly reported system Yb$_3$Pt$_4$~\cite{kobayashi.2007,aronson.2008}. 
In the paramagnetic phase Yb$_3$Pt$_4$ has the Curie-like susceptibility 
and nearly constant resistivity which exceeds the spin-disorder limit. 
Like in  YbInCu$_4$, the LM phase of  Yb$_3$Pt$_4$ looks semimetallic, 
with unrenormalized ('light') Bloch sates and localized $f$ states. 
In the absence of the $f$--$c$ exchange coupling, the paramagnetic 
entropy cannot be reduced by the Kondo screening but the ground 
state has to be approached by an alternative route.
In Yb$_3$Pt$_4$  the ground state is magnetic and we speculate that 
the LRO is due to a spin density wave (SDW). The transition between 
the LM and the SDW phase takes place at $T_N\simeq 2$ K.   
The data show that the effective mass of metallic electrons becomes 
heavy at $T_N$~\cite{aronson.2008}. 

This behavior can be explained by the slave boson approach,  
assuming that the chemical potential is close to the pseudo 
gap of the unperturbed $c$ band and admitting an AFM solution.
The generalized slave boson equations which include the coupling 
to the staggered magnetization provide at $T_N$ a non-trivial solution
with hybridized states and a large FS.
The reason is that large internal fields shift the chemical 
potential of up and down spins away from the singularity in $c$ DOS  
(like in the case  of an external magnetic field, discussed in 
section~\ref{B}). Since the Kondo coupling in the LM phase 
is very small, the $c$ electrons are nearly free and 
the Kondo screening can only occur at extremely low temperatures. 
Unlike YbInCu$_4$, where the paramagnetic entropy is removed 
by an iso-structural phase transition which enables hybridization 
and the transition to the FL phase,  
in Yb$_3$Pt$_4$ the paramagnetic entropy is removed by 
a spin density wave (SDW) transition which partially 
gaps the Fermi surface. 
The formation of the SDW at temperature $T_N\simeq$ 2 K switches-on 
the hybridization, which delocalizes the $f$ states, changes the 
effective degeneracy of the $f$ states, and renormalizes the metallic mass.
The chemical potential in the SDW phase is still rather close to 
the pseudo-gap, so that the ensuing FL scale is 
large, $T_0 \gg T_N$ (see discussion in Sec.~\ref{T0}). 
Hence, the mass enhancement at the transition is moderate or small.
It would be interesting to check the assumption about the pseudogap 
in the $c$ DOS by performing the band structure calculations that would 
take into account the large Coulomb repulsion between the $f$ electrons.

\subsubsection{The  $T_0 \simeq T_K $ case  \label{T0=TK} } 

As a final example we consider the anomalous pressure dependence 
of the $T^2$ coefficient of the resistivity, $A(P)$, 
and the residual resistance, $\rho_0(P)$, observed in 
heavy fermions like CeCu$_2$Si$_2$\cite{holmes.04}, 
CeCu$_2$Ge$_2$\cite{jaccard.99} or CePd$_2$Ge$_2$\cite{wilhelm.02} 
at very low temperatures. 
In these compounds, the FL scale is about the same as the Kondo 
scale inferred from the low-temperature peak in $\rho(T)$ 
or $\alpha(T)$\cite{jaccard.99,zlatic.08}. The scales T$_0$ and T$_K$ 
are much smaller than the CF splitting (estimated from the high-temperature 
peak in in $\rho(T)$ or $\alpha(T)$\cite{zlatic.08}), 
so that the excited CF sates can be neglected for $T\leq$T$_K\ll\Delta_{CF}$. 
At ambient pressure, the ground state of these compounds is often 
superconducting or magnetic and the FL behavior is enforced 
by applying an initial pressure $P_0$. 
For $P > P_0$, the data show that $A(P)$ decreases gradually 
from large initial values, drops at a critical pressure $P_c$ 
by nearly two orders of magnitude\cite{jaccard.99,wilhelm.02,holmes.04}, 
and then continues a gradual decrease. 
The residual resistance increases from a small initial value, 
rises rapidly to a sharp peak at $P_c$, and decreases 
at very high pressure to rather small values\cite{jaccard.99,holmes.04}. 

We explain these features by assuming that for 
$P_0\leq P \leq P_c$ the effective degeneracy of the model 
is defined by the lowest CF state and that pressure changes 
the $f$--$c$ coupling and T$_K$ but not $\Delta_{CF}$.
The FL state established at $P_0$ has a large Fermi surface, 
because the $f$ electrons are delocalized and participating 
in the Fermi sea.
Taking,  for simplicity, the half-filled conduction band and  
describing the 4$f$ state be a CF doublet, 
we find that the Fermi surface is close to the 
edge of the Brillouin zone and that the average Fermi velocity 
in Eq.~(\ref{eq: rho_gama}) is small. 
In the aforementioned compounds the FL scale is small, 
so that $A(P_0)$ large. 
As long as pressure does not change the degeneracy of the 
$f$ states, Luttinger theorem preserves the Fermi surface, 
so that $v_F^2$ and ${{\Delta\mu}}$ in Eq.~(\ref{eq: rho_gama} ) 
remain approximately constant. 
Since $N$ is also constant in this pressure range, the pre-factor 
of the $(\gamma T)^2$-term in Eq.~(\ref{eq: rho_gama}) does not 
change much. Thus, the main effect of pressure is a gradual 
reduction of $A(P)$ from the maximum value attained at $P_0$. 
This reduction is due to the increase of $T_0(P)$, as can be seen 
from the linear scaling between $\sqrt{A(P)}$ and the inverse 
Kondo scale of the system $1/$T$_K(P)$\cite{jaccard.99}.  

At large enough pressure the hybridization becomes too large for 
the system to support the CF excitations, so that the effective 
degeneracy of the ground state increases. 
For $P \geq P_c$, the degeneracy is not set by the lowest CF level 
but by the full multiplet (a sextet, in the case of Cerium).
The slave boson solution of the periodic Anderson model 
with the $SU(N)$ symmetry and infinite correlation shows that 
the Fermi volume decreases as $N$ increases, because a single $f$ 
electron has to be distributed over more and more channels. 
The chemical shift ${{\Delta\mu}}$ and the specific heat 
coefficient $\gamma$ also decrease with $N$, while $v^2_F$ becomes 
larger on a smaller FS, so that $A(P)$ drops sharply for $P\geq P_c$. 
The degeneracy of the $f$ state does not change for $P > P_c$ 
and an increase of pressure above $P_c$ reduces $A(P)$ by 
increasing T$_0$. 
In this pressure range, we find the scaling between ${A(P)}$ 
and $1/$T$_K^2$, where T$_K^2$ is the Kondo scale of the $N$-fold 
degenerate model. 

The residual resistance also tracks the pressure-induced changes 
in the effective degeneracy. If the $f$ states are localized 
at ambient pressure, as seems to be the case with 
CeCu$_2$Ge$_2$\cite{jaccard.99} or CePd$_2$Ge$_2$\cite{wilhelm.02} 
at very low temperatures, the initial values of $\rho_0(P_0)$ are small.  
Taking, for simplicity, the model with a ground state doublet and an 
excited CF quartet, and using the FL laws, we find that the temperature 
dependence of the resistivity at ambient pressure is due to two 
resonant channels, while most of the current is carried by four 
non-resonant channels which have temperature-independent conductivity. 
The contribution of the resonant channels to the residual 
resistance can be neglected. 
At large enough pressure, $P\simeq P_c$, the CF excitations are removed, 
the degeneracy of the ground state changes from doublet to sextet,  
and all the channels become resonant. 
Because of Luttinger theorem, the Fermi volume shrinks 
for two of the channels (former resonant ones) and expands 
for four of them  (former non-resonant ones). 
Hence,  for $P\simeq P_c$, the overall contribution to 
the residual resistance increases sharply.  A further increase of 
pressure does not change the degeneracy of the model but 
gives rise to the charge transfer from the $f$ to the $c$ states.  
This reduces the residual resistance and gives the $\rho_0(P)$ 
curve its asymmetric shape.  
In other words, the FL liquid laws and the slave boson solution 
of the periodic Anderson model show that the large peaks 
in $A(P)$ and $\rho_0(P)$ are due to the pressure-induced 
change in the degeneracy of the $f$ state.

\section{Summary and conclusions} 
It is well known that the degeneracies and the splittings 
of the 4$f$ states have a strong impact on the Kondo scale 
and the high-temperature behavior of intermetallic compounds 
with Ce, Eu or Yb ions. 
In this paper, we have shown that the details of the conduction 
electron band structure have a large impact on the ratio 
of the Kondo scale to the Fermi liquid scale, 
which determines the type of the crossover between 
the incoherent and coherent regimes. 

Our analysis is based on the periodic Anderson model, which 
can be mapped at high temperatures on a single impurity 
Anderson or Kondo models with Kondo scale T$_K$. 
The incoherent properties of the lattice can be understood 
using a single impurity approximation which takes into account 
the structure of the $f$ states. 
At low temperatures, the slave boson solution of the periodic model 
yields the FL laws characterized by an energy scale T$_0$. 
The Kondo and the FL scales depend on the shape of 
the $c$ DOS in the vicinity of the chemical potential, 
the degeneracy and the CF splitting of the $f$ states, 
the number of $c$ and $f$ electrons, and their coupling. 
The ratio T$_0/$T$_K$ is determined by the details of the 
band structure. 
Depending on the relative magnitude of T$_0$ and T$_K$, 
the crossover between the high- and low-temperature regimes 
proceeds along very different routes. 
A sharp peak in the $c$ DOS yields T$_0 \ll$T$_K$ and gives  
rise to a 'slow crossover', as observed in YbAl$_3$ and similar compounds. 
The minimum in the $c$ DOS yields T$_0 \gg$T$_K$, which causes  
an abrupt transition between the high- and low-temperature regimes,  
as observed in YbInCu$_4$-like systems. 
In the case of CeCu$_2$Ge$_2$ and CeCu$_2$Si$_2$, where  
T$_0 \simeq$T$_K$, our results show that the pressure-dependence 
of the $A(P)$ coefficient and the residual resistance can be 
related to the change in the degeneracy of the $f$ states. 

The slave boson solution shows that the low-temperature 
response of the periodic model can be enhanced (or reduced) 
with respect to the predictions of the single-impurity model 
with the same Kondo scale. 
In the coherent regime, the renormalization of transport coefficients 
modifies the Wiedemann-Franz law and can lead to an enhancement 
of the thermoelectric figure-of-merit. 
The FL laws explain the correlation between the specific heat 
coefficient $\gamma$ and the slope of the thermopower, $\alpha(T)/T$, 
or between $\gamma$ and the T$^2$ coefficient of the electrical 
resistance $A=\rho(T)/T^2$. 
In the case of a $N$-fold degenerate model, the FL laws explain 
the deviations of the Kadowaki-Woods ratio $R_{KW}=A/\gamma^2$ and 
the $q$ ratio, $q=|e|lim_{\{T\to 0\}x} \alpha/\gamma T$, from 
the universal values. 

The shape of the $c$ DOS affects also the magnetic response of 
the system. The field-dependence of the magnetization resembles 
the temperature-dependence of the entropy, as can be seen from 
Figs.~\ref{SchemasMagnetization} and  ~\ref{Schemasentropy}.  
The $c$ DOS with a maximum around the chemical potential 
leads to an extended plateau in $m(h)$, while a pseudo-gap 
in the $c$ DOS leads to a metamagnetic transition for the fields 
of the order of the Kondo temperature.

\acknowledgments
The authors are grateful to R. Monnier, J. Freericks and M. Vojta 
for useful discussions. 
S.B. acknowledges the Max Planck Institute for the Physics of 
Complex Systems, Dresden, Germany, where part of this work has 
been done. Support from the Ministry of Science of Croatia under 
Grant No. 0035-0352843-2849, the NSF under Grant No. DMR-0210717, 
and the COST P-16 ECOM project is acknowledged.

\begin{appendix} 
 
\section{Generalization to multi-orbital systems} 
\subsection{General formalism} 
Here, we generalize our model to more realistic systems where the local f  
orbital have a supplementary degeneracy. The later is lift at low temperature  
by the crystal field. In the limit $U\to\infty$, the PAM  
Hamiltonian~(\ref{PAM}) is generalized to  
\begin{eqnarray} 
H=\sum_{{\bf k}\sigma} 
\epsilon_{\bf k}c_{{\bf k}\sigma}^{\dagger}c_{{\bf k}\sigma} 
+\sum_{i\alpha\sigma} 
(E_{f}+\Delta_{CF}^{\alpha}) 
f_{i\alpha\sigma}^{\dagger}f_{i\alpha\sigma} 
+V\sum_{i\alpha\sigma}[c_{i\sigma}^{\dagger}f_{i\alpha\sigma} 
+f_{i\alpha\sigma}^{\dagger}c_{i\sigma}] 
-\mu\sum_{i\sigma}\left[
\sum_{\alpha}f_{i\alpha\sigma}^{\dagger}f_{i\alpha\sigma} 
+c_{i\sigma}^{\dagger}c_{i\sigma}\right]~,  
\end{eqnarray} 
where $\alpha=1,\cdots, N/2$ is a local orbital index, and  
$\Delta_{CF}^{\alpha}$ is the crystal field.  
For Ce and Yb we have $N=6$ and $N=8$.  
The slave boson approach described in section~\ref{formalism} 
for $N=2$ can be easily generalized  
within the mapping $\left| 0 \right>_{i}^{f} 
\longrightarrow  
b_{i}^{\dagger}\left| 0 \right>_{i}$,  
$\left| \alpha\sigma \right>_{i}^{f} 
\longrightarrow  
d_{i\alpha\sigma}^{\dagger}\left| 0 \right>_{i}$,  
with the exclusion of all the states with more than one electron on a  
given site.  
This leads to the local identities  
$b_{i}^{\dagger}b_{i}+ 
\sum_{\alpha\sigma}f_{i\alpha\sigma}^{\dagger}f_{i\alpha\sigma}=1$,  
which are satisfied by introducing Lagrange multipliers $\lambda_{i}$.  
Within the mean-field approximation, we replace the bosonic fields $b_{i}$  
and the Lagrange multipliers $\lambda_{i}$  
by homogeneous static real fields $r$, and $\lambda$.  
The quadratic mean-field Hamiltonian~(\ref{PAMmagnetbosonsMF}) 
obtained for $N=2$  
is then generalized to  
\begin{eqnarray} 
H=\sum_{{\bf k} \sigma}\left[ { 
\epsilon_{\bf k}c_{{\bf k}\sigma}^{\dagger}c_{{\bf k}\sigma} 
+rV\sum_{\alpha}\left( { 
c_{{\bf k}\sigma}^{\dagger}d_{{\bf k}\alpha\sigma} 
+d_{{\bf k}\alpha\sigma}^{\dagger}c_{{\bf k}\sigma} 
}\right) 
-\mu c_{{\bf k}\sigma}^{\dagger}c_{{\bf k}\sigma} 
+ 
\sum_{\alpha}(\lambda+\Delta_{CF}^{\alpha}) 
d_{{\bf k}\alpha\sigma}^{\dagger}d_{{\bf k}\alpha\sigma} 
+\frac{E_{f}-\lambda-\mu}{2}(1-r^{2}) 
}\right]~.   
\label{PAMCFbosonsMF} 
\end{eqnarray} 
The self-consistent parameters $r$ and $\lambda$ are obtained  
by minimizing the free energy  
$\beta {\cal F}(r, \lambda)\equiv -\ln Tr e^{-\beta H}$,   
such that ${\partial {\cal F}(r, \lambda)}/{\partial r}=0$ 
and ${\partial {\cal F}(r, \lambda)}/{\partial \lambda}=0$.  
This gives  
\begin{eqnarray} 
2r (E_{f}-\lambda-\mu) 
&=& 
V\sum_{{\bf k}\alpha\sigma}\langle c_{{\bf k}\sigma}^{\dagger}d_{{\bf k}\alpha\sigma} 
+d_{{\bf k}\alpha\sigma}^{\dagger}c_{{\bf k}\sigma}\rangle~,   
\label{MFCFeqr1}\\ 
r^{2} 
&=& 
1-\sum_{{\bf k}\alpha\sigma} 
\langle d_{{\bf k}\alpha\sigma}^{\dagger}d_{{\bf k}\alpha\sigma}\rangle~,   
\label{MFCFeqlambda1} 
\end{eqnarray} 
where, $\langle\cdots\rangle$ denotes the thermal average with respect to  
the mean-field Hamiltonian~(\ref{PAMCFbosonsMF}) 
and  
\begin{eqnarray}  
n=\sum_{i\sigma}\langle  
\sum_{\alpha}d_{i\alpha\sigma}^{\dagger}d_{i\alpha\sigma} 
+c_{i\sigma}^{\dagger}c_{i\sigma}\rangle 
\equiv \sum_{\alpha}n_{f}^{\alpha}+n_{c}~.  
\label{MFCFeqmu1} 
\end{eqnarray} 
The average electronic occupation per site of the $f$ and $c$ orbitals is   
$n_{f}^{\alpha}\equiv \sum_{i\sigma}\langle f_{i\alpha\sigma}^{\dagger}f_{i\alpha\sigma}\rangle 
=\sum_{i\sigma}\langle d_{i\alpha\sigma}^{\dagger}d_{i\alpha\sigma}\rangle$  
and  
$n_{c}\equiv \sum_{i\sigma}\langle c_{i\sigma}^{\dagger}c_{i\sigma}\rangle$,   
respectively.  
These averages are determined by the slave-boson amplitude and  
the total electronic occupation,   
\begin{eqnarray} 
               \label{eq: CFnf} 
\sum_{\alpha}n_{f}^{\alpha}&=&1-r^{2}~, \\ 
n_{c}&=&n-\sum_{\alpha}n_{f}^{\alpha}~.  
\end{eqnarray} 
The self-consistent  
Eqs.~(\ref{MFCFeqr1}, \ref{MFCFeqlambda1}, \ref{MFCFeqmu1})  
can be rewriten as  
\begin{eqnarray} 
r\frac{E_{f}-\lambda-\mu}{2} 
&=& 
-V\int_{-\infty}^{+\infty}\rho_{dc}(\omega)n_{F}(\omega)d\omega 
\label{MFCFeqr2}~, \\ 
\frac{n_{f}^{\alpha}}{2} 
&=& 
\int_{-\infty}^{+\infty}\rho_{d}^{\alpha}(\omega)n_{F}(\omega)d\omega 
\label{MFCFeqlambda2}~, \\ 
\frac{n_{c}}{2} 
&=& 
\int_{-\infty}^{+\infty}\rho_{c}(\omega)n_{F}(\omega)d\omega 
\label{MFCFeqmu2}~,  
\end{eqnarray} 
where $n_{F}(\omega)\equiv 1/[1+e^{\beta\omega}]$ is the Fermi function 
and $\rho_{dc}$, $\rho_{d}^{\alpha}$ and $\rho_{c}$  
are the spectral densities of  
the local single-particle Green's functions of a given spin. 
These Green's functions are defined in the usual way as   
thermal averages of the (imaginary) time-ordered products of the  
appropriate creation and annihilation operators.  
\begin{eqnarray} 
\rho_{dc}(\omega)&\equiv& 
-\frac{1}{\pi} \sum_{{\bf k}}\sum_{\alpha} 
{\mathrm Im} \ G_{dc}^{\alpha}({\bf k},\omega)~, \\ 
\rho_{d}^{\alpha}(\omega)&\equiv& 
-\frac{1}{\pi} \sum_{{\bf k}} 
{\mathrm Im} \ G_{d}^{\alpha}({\bf k},\omega)~, \\ 
\rho_{c}(\omega)&\equiv& 
-\frac{1}{\pi} \sum_{{\bf k}} 
{\mathrm Im} \ G_{c}({\bf k},\omega)~.  
\end{eqnarray} 
The Green functions are computed from the quadratic  
Hamiltonian~(\ref{PAMCFbosonsMF}). We find  
\begin{eqnarray} 
G_{cc}({\bf k},\omega ) 
&=& 
\frac{1}{\omega+\mu-\epsilon_{\bf k} 
-r^{2}V^{2}\sum_{\alpha}1/(\omega-\lambda-\Delta_{CF}^{\alpha})}~, \\ 
G_{dc}^{\alpha}({\bf k},\omega ) 
&=& 
-\frac{rV}{\omega-\lambda-\Delta_{CF}^{\alpha}}G_{cc}({\bf k},\omega )~, \\ 
G_{dd}^{\alpha}({\bf k},\omega ) 
&=& 
\frac{1}{\omega-\lambda-\Delta_{CF}^{\alpha}} 
+ 
\frac{r^{2}V^{2}}{(\omega-\lambda-\Delta_{CF}^{\alpha})^{2}} 
G_{cc}({\bf k},\omega )~.  
\end{eqnarray} 
We define the non-interacting electron Green's function  
\begin{eqnarray} 
G_{0}(\omega ) 
\equiv  
\sum_{\bf k} \frac{1}{\omega-\epsilon_{\bf k}}~.  
\end{eqnarray} 
The local spectral densities can be rewritten as  
\begin{eqnarray} 
\rho_{c}(\omega)&=&-\frac{1}{\pi} {\mathrm Im} \ G_{0}\left(  
\omega+\mu-r^{2}V^{2}\sum_{\alpha} 
\frac{1}{\omega-\lambda-\Delta_{CF}^{\alpha}} 
\right)~, \\ 
\rho_{dc}^{\alpha}(\omega)&=&\frac{1}{\pi} {\mathrm Im} \ \left[  
\frac{rV}{\omega-\lambda-\Delta_{CF}^{\alpha}} 
G_{0}\left(  
\omega+\mu-r^{2}V^{2}\sum_{\alpha} 
\frac{1}{\omega-\lambda-\Delta_{CF}^{\alpha}} 
\right) 
\right]~, \\ 
\rho_{d}^{\alpha}(\omega)&=&-\frac{1}{\pi} {\mathrm Im} \ \left[  
\frac{1}{\omega-\lambda-\Delta_{CF}^{\alpha}} 
+ 
\frac{r^{2}V^{2}}{(\omega-\lambda-\Delta_{CF}^{\alpha})^{2}} 
G_{0}\left(  
\omega+\mu-r^{2}V^{2}\sum_{\alpha} 
\frac{1}{\omega-\lambda-\Delta_{CF}^{\alpha}} 
\right) 
\right]~.  
\end{eqnarray} 
 
\subsection{Degenerate case: no crystal field splitting} 
We consider here $\Delta_{CF}^{\alpha}=0$.  
In this case, we can check easily that the complete formalism described in  
this article for $N=2$ can be generalized to any value of $N$, within the  
following rescaling 
\begin{eqnarray} 
V^{2} \longrightarrow \frac{N}{2}V^{2}~.  
\end{eqnarray} 
Note that, due to the exponential dependence of the energy scales  
$T_{0}$ and T$_{K}$ with respect to $V^{2}$  
(see Eqs.~(\ref{DefKondocoupling}-\ref{ExpressionTK}-\ref{ExpressionTzero}),  
these two scales have to be rescaled like  
\begin{eqnarray} 
T_{K}^{(N)}=T_{K}^{(2)}e^{-(2-N)/NJ(\mu_{0})\rho_{0}(\mu_{0})}~, \\ 
T_{0}^{(N)}=T_{0}^{(2)}e^{-(2-N)/NJ(\mu_{0})\rho_{0}(\mu_{0})}~, 
\end{eqnarray} 
where $(N)$ refers to the number of degenerate channels considered  
for the $f$ orbitals.  
 
\subsection{Kondo temperature} 
For $r\to 0$, the generalized mean-field equations can be written as  
\begin{eqnarray} 
\frac{n_{c}}{2} 
&=& 
\int_{-\infty}^{+\infty}\rho_{0}(\omega+\mu)n_{F}(\omega )d\omega~, \\ 
\frac{n_{c}^{\alpha}}{2} 
&=& 
n_{F}(\lambda+\Delta_{CF}^{\alpha})~, \\ 
\frac{E_{f}-\lambda-\mu}{V^{2}} 
&=&\sum_{\alpha=1}^{N/2} 
\int_{-\infty}^{+\infty} 
\frac{1}{\omega-\lambda-\Delta_{CF}^{\alpha}} 
\rho_{0}(\omega+\mu)\tanh\left( \frac{\omega}{2T}\right) 
d\omega~. 
\end{eqnarray} 
We can arbitrarily choose one vanishing  
$\Delta_{CF}^{\alpha=0}=0$, corresponding to the doublet with the lowest  
local energy $E_{f}$. If, for the other doublets,  
one has $\Delta_{CF}^{\alpha\neq 0}>>T$, then we obtain  
$n_{f}^{\alpha\neq 0}=0$, and the formalism developed for $N=2$ can  
be applied simply.  
In the more general case where $\Delta_{CF}^{\alpha\neq 0}\sim T$,  
or $\Delta_{CF}^{\alpha\neq 0}< T$ then  
one has to consider also the contribution from these orbitals.

\end{appendix} 

\bibliographystyle{apsrev}
\bibliography{sb_pam}
\end{document}